\documentclass[a4paper,10pt,preprint,5p]{elsarticle}

\usepackage[breaklinks]{hyperref}
\usepackage{breakurl}
\usepackage{amsmath}

\newcommand{\e}{e}
\newcommand{\bra}[1]{\ensuremath{\langle #1 |}}
\newcommand{\ket}[1]{\ensuremath{|#1\rangle}}

\def\lsim{\mathrel{\rlap{\lower4pt\hbox{\hskip0pt$\sim$}}
    \raise1pt\hbox{$<$}}}
\def\gsim{\mathrel{\rlap{\lower4pt\hbox{\hskip0pt$\sim$}}
    \raise1pt\hbox{$>$}}}

\begin{document}

\begin{frontmatter}

\title{Continuous variables quantum computation over the vibrational modes of a single trapped ion}

\author[mymainaddress]{Luis Ortiz-Guti\'e{}rrez\tnoteref{mycorrespondingauthor}}
\cortext[mycorrespondingauthor]{Corresponding author}
\ead{lortiz@df.ufpe.br}
\author[mymainaddress]{Bruna Gabrielly}
\author[mymainaddress]{Luis F. Mu\~n{}oz}
\author[mymainaddress]{Kain\~a T. Pereira}
\author[mysecondaryaddress]{Jefferson G. Filgueiras}
\author[mythirdaddress]{Alessandro S. Villar}

\address[mymainaddress]{Departamento de F\'\i{}sica, Universidade Federal de Pernambuco, 50670-901 Recife, PE, Brazil}
\address[mysecondaryaddress]{Instituto de F\'\i{}sica de S\~ao Carlos, Universidade de S\~ao Paulo, P.O. Box 369, S\~ao Carlos, 13560-970 SP, Brazil}
\address[mythirdaddress]{American Physical Society, 1 Research Road, Ridge, New York 11961, USA}

\begin{abstract}
We consider the quantum processor based on a chain of trapped ions to propose an architecture wherein the motional degrees of freedom of trapped ions (position and momentum) could be exploited as the computational Hilbert space. We adopt a continuous-variables approach to develop a toolbox of quantum operations to manipulate one or two vibrational modes at a time. Together with the intrinsic non-linearity of the qubit degree of freedom, employed to mediate the interaction between modes, arbitrary manipulation and readout of the ionic wave function could be achieved.
\end{abstract}


\end{frontmatter}


\section{Introduction}

The current paradigm of implementations of quantum computing consists in the coherent manipulation of discrete two-level systems, the qubits, by sequences of quantum gates~\cite{nielsenchuangQCQI}. Ion traps stand as one of the most successful experimental implementations of a quantum processor, a system for which all basic elements required for computation universality have been demonstrated in proof-of-principle experiments~\cite{winelandCNOT_prl95,phasegate_nature03,cnot_nature03}  and scalability is not unlikely~\cite{CZscalable_nature00,scalabletrap_nature02}.

A chain of ions trapped in a harmonic potential functions as a quantum register wherein each ion encodes one qubit in energy eigenstates of its electronic configuration~\cite{ciraczoller95}. The external confinement and the electric repulsion among ions give rise to collective modes of vibration which are employed to mediate the interaction between any chosen pair of qubits. Quantum operations are accomplished by resonant or near-resonant laser pulses with the qubit transitions. Other internal energy levels of the ions are employed to initialize and measure the qubits. Even though the specifics of this manipulation scheme has evolved enormously since its inception~\cite{ms_prl99,sigmaz_00,phasegate_nature03,ultrafastentanglementmonroe_prl13,spindependentMonroe_05}, it would not be inappropriate to name it as the `Cirac \& Zoller (CZ) paradigm' of ion trap quantum computing. In short, the CZ paradigm has each ion storing a single qubit in internal electronic energy levels and different qubits interacting via the quantum information `bus' provided by one or more motional modes.

An alternative route to quantum computation considers physical observables with continuous spectra -- continuous variables (CV) -- to realize the physical encoding and manipulation of quantum information~\cite{lloydbraunstein_prl99,reviewbraunsteinvanloock_rmp05}. 
In the continuous variables quantum computing (CVQC) paradigm, Gaussian states and operations are usually considered as the building blocks of quantum logic~\cite{symplectic,eisert_ijqi03}, as well as a single non-Gaussian operation needed to achieve universality~\cite{bartlett_prl02,bartlett_pra02}. The basic physical object of quantum computing is embedded in this case in an infinite-dimensional Hilbert space. And although it may be regarded as continuous in the eigenbasis of certain observables, it can many times also be understood as a discrete configuration space in the eigenbasis of other observables. More concretely, as considered in this paper, a vibrational mode of the ion chain~\cite{wineland_science96,zagury_pra99,wineland2_prl96,mesoscopicmotion_prl01,qbitenconding100photon_science13} can be either described in the continuous phase space of position and momentum observables, e.g. by the Wigner function, or in terms of superpositions of the quantized energy eigenstates of the harmonic oscillator, the number or Fock states~\cite{haroche_exploringthequantum}. The generation, control~\cite{Alessio2009,Alonso2013,Ding2014} and measurement~\cite{Mirkhalaf2012,Johnson2015,Hashemloo2016} of vibrational modes have been approached in the recent literature in various ways: by entangling them with optical resonator modes~\cite{Nicacio2013}, using them as equivalent models for the study of vibrational states of optomechanical systems~\cite{Xu2013b}, by generating exotic quantum states~\cite{Miry2012,Roriguez2012}, or  by implementing quantum simulations of solid state systems~\cite{Haze2012,Dutta2012,Dutta2013}. 

In this paper, we investigate the idea of exploiting the vibrational modes of trapped ions as the physical platform of quantum computing, i.e. for the implementation of quantum gates in the motional modes of vibration~\cite{Orszag2005,Zhang2008}. We consider the feasibility and particularities of inverting the CZ paradigm to employ the qubit degree of freedom as the mediator of interaction among a set of motional modes of vibration. Even though our proposal can be extended to a system of different singly trapped ions~\cite{Lau2012}, we focus here on the simplest case of a single trapped ion and its corresponding set of three vibrational modes as a starting point. By following this approach, we try to establish the potential capabilities brought by this minimalistic quantum system and the likely limitations on the size of the configuration space made available by this simple change of perspective in the use of the ion trap. We develop a CV quantum computation toolbox to manipulate each of the single modes and to make them interact in pairs, in particular to show that conditional dynamics (entangling gates) would be available. The proposed quantum gates are realizable by bichromatic laser fields with tunable frequencies. Readout of the quantum state can be performed using number-dependent Rabi flops on the qubit~\cite{matosfilhovogel_prl96,davidovich_pra96,lutterbachdavidovich_prl97}.

This paper is organized as follows. In Sec.~\ref{sec:basichamiltonian}, we present the quantum processor based on the ion trap and recall the basic manipulation of a single trapped ion by an external laser source. Sec.~\ref{sec:bichromatic} presents the CV quantum gates that can be realized in the motional modes with bichromatic laser fields. We detail our CVQC proposal and develop the necessary toolbox of quantum gates in Sec.~\ref{sec:toolbox}. Our concluding remarks follow in Sec.~\ref{sec:conclusion}.

\section{Basic implementation}
\label{sec:basichamiltonian}

\subsection{Physical system}

In this proposal, the physical objects to be manipulated are the different oscillation modes of a quantum harmonic oscillator. There are three available modes in the simplest case of a single trapped ion oscillator. The size of the Hilbert space associated with each vibrational mode and available to manipulation in actual experimental conditions is better quantified in the eigenbasis of the number operators. The basis for each mode is assumed to be truncated at a maximum phonon number $N$, and is hence composed of the eigenstates
\begin{equation}
\label{eq:quantumstatefockbasis}
\{\ket{0},\ket{1},\ket{2},\dots,\ket{N}\}.
\end{equation}

\begin{figure}[htbp]
\centerline{\includegraphics[width=0.95\columnwidth]{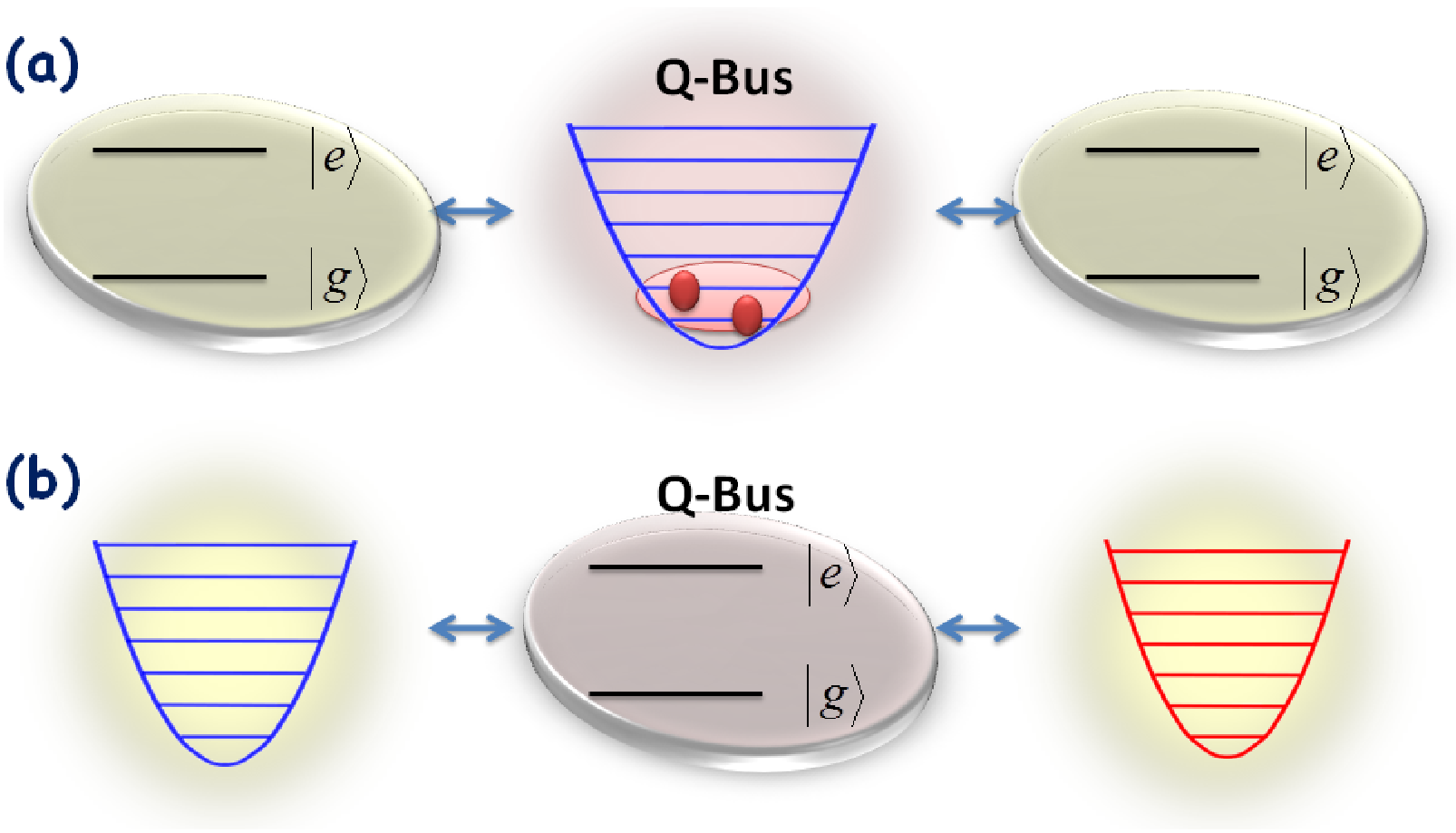}}
\caption{Proposal of CVQC using the ion trap. (a) The CZ paradigm employs one or more motional modes of the quantum harmonic oscillator to mediate the interaction (the `quantum bus') between pairs of qubits. In the original proposal, one motional mode is essentially employed as a two-level system. Increasing the size of the Hilbert space requires adding more ions to the physical system, reaching an exponential increase. (b) Our proposal utilizes the qubit as the quantum bus that allows vibrational modes to interact. Three harmonic oscillator form the Hilbert space available for manipulation. The size of the Hilbert space available is outright larger than in the CZ paradigm, although it increases only polynomially with the phonon cap number. }
\label{fig:schemeproposal}
\end{figure}

The representation of the quantum state in terms of phonon number eigenstates refers to a `particle-like' description of the quantum system. The quantum state of the harmonic oscillator also admits a CV representation in the position and momentum phase space, a `wave-like' description employing the Wigner function. 
In phase space, quantum gates are transformations of the Wigner function. In the ion trap processor, they can be performed by coupling vibrational modes to the qubit (internal) degree of freedom by means of bichromatic laser light~\cite{ms_prl99,sigmaz_00,ms_pra00,roos_njp08}. The qubit is by construction a highly non-linear physical system -- one that saturates with a single quantum --, a property here employed to generate non-Gaussian operations on the vibrational modes. Since in our proposed ion trap CVQC architecture the qubit is only an auxiliary source of non-linearity and coupling among motional modes, the desired quantum operations must start and end with quantum states $\hat\rho$ which are separable in the qubit $\hat\rho_{q}$ and motional modes $\hat\rho_m$, i.e. we impose that $\hat\rho=\hat\rho_{q}\hat\rho_m$ before and after the application of quantum gates. 

The CV toolbox of quantum operations to be developed below can be separated in Gaussian and non-Gaussian operations. The class of Gaussian operations maintains as Gaussian an initially Gaussian Wigner function. There are single- and two-mode Gaussian operations. Single-mode displacements and squeezers respectively displace the origin of phase space or the scaling of the position and momentum axis. Both of them have already been experimentally demonstrated in the ion trap processor~\cite{wineland_prl96,squeezingHome_nature15}. Two-mode operations comprise the beam splitter and the two-mode squeezer. The beam splitter is a passive transformation that linearly combines two field modes. The two-mode squeezer, an active transformation, can be understood as two single-mode squeezers simultaneously acting on orthogonal combinations of two modes. One can also include two-mode conditional gates as generalizations of such operations.

\subsection{ Hamiltonian of the ion trap}

Our CVQC toolbox is built upon the simplest implementation of an ion trap processor: a single ion furnishes the qubit and three independent modes of vibration. We consider in this section the basic coherent manipulation of a single trapped ion by an external source of coherent light~\cite{rmpiontrap_03,haroche_exploringthequantum}.

To establish notation, we recall below the elementary dynamics of one qubit and two motional modes coupled to it by a monochromatic external laser. The generalization of the interaction to three oscillator modes and bichromatic lasers capable of producing the desired quantum gates follows next.

The ion trap Hamiltonian reads in this case as
\begin{equation}
\hat H = \hat H_0 + \hat H_I,
\end{equation}
where $\hat H_I$ is the interaction Hamiltonian discussed below and $\hat H_0$ provides the free dynamics of qubit and motional modes,
\begin{equation}
\hat H_0 = {\textstyle\frac12}\hbar \omega_0\hat\sigma_z + \hbar \omega_a \hat a^\dag \hat a + \hbar \omega_{b} \hat b^\dag \hat b.
\end{equation}
The qubit transition frequency is $\omega_0$ and its two-dimensional Hilbert space is described in terms of the excited $\ket{e}$ and ground $\ket{g}$ internal states of the ion, with which we write $\hat \sigma_z = \ket{e}\bra{e}-\ket{g}\bra{g}$. The two independent vibrational modes under consideration are described in terms of the annihilation operators $\hat a$ and $\hat b$ and the respective creation operators satisfying the commutation relations $[\hat a,\hat a^\dag]=[\hat b,\hat b^\dag]=1$. Their oscillation frequencies are $\omega_s$, where $s\in\{a,b\}$ denotes the mode.

The simplest model of interaction Hamiltonian $\hat H_I=-\vec d\cdot \vec E$ comprises a dipolar coupling between the ion and an external coherent light source. The atomic dipole operator is $\vec d = \vec\mu (\hat\sigma_+ +\hat\sigma_-)$, with dipole moment $\vec\mu$ and operators $\hat\sigma_+=\ket{e}\bra{g}$ and $\hat\sigma_-=\ket{g}\bra{e}$. The light source drives the ion by means of the electric field $\vec E = \vec E_0 \exp(i \vec k\cdot \vec r - i\omega_L t)$ with wavevector $\vec k$ and frequency $\omega_L$. The interaction Hamiltonian can be made to account for the free evolution associated with $\hat H_0$ (interaction picture), yielding
\begin{align}
\tilde H_I = & \;{\textstyle\frac12}\hbar\Omega \hat\sigma_+ \e^{-i\delta t}  \exp[i\eta_a (\hat a e^{-i\omega_a t} + \hat a^\dagger e^{i\omega_a t}) \\
& \qquad \qquad\qquad\quad + i\eta_b (\hat b e^{-i\omega_b t} + \hat b^\dagger e^{i\omega_b t})] + \mathrm{h.c.},\nonumber
\end{align}
where $\delta = \omega_L-\omega_0$ is the radiation-atom detuning, $\Omega=|\vec\mu\cdot\vec d|/\hbar$ is the Rabi frequency, and $\eta_s=kx_s\cos\theta$ are the Lamb-Dicke parameters, defined in terms of the typical scale of the ground state oscillator wavefunction $x_{s} = \sqrt{\hbar/(2m\omega_{s})}$ and the direction of propagation $\theta$ of the laser with respect to the direction of vibration of mode $s$. Typical experimental conditions in optical qubits imply $\eta_{s}\ll 1$, values for which the Lamb-Dicke regime can be evoked to expand the interaction Hamiltonian in powers of $\eta_{s}$.

The CV quantum gates we consider in the next section are obtained by expanding the interaction Hamiltonian up to second order in $\eta_s$~\cite{Roghani,Moya}, as
\begin{align}
\label{eq:Hexpansion}
\tilde H_I = & \; \hat H^{(0)} + \eta_a \hat H_a^{(1)} + \eta_b \hat H_b^{(1)} \\
& - \eta_a^2 \hat H_{a}^{(2)} - \eta_b^2 \hat H_{b}^{(2)} - 2\eta_a\eta_b \hat H_{ab}^{(2)} + \mathcal{O}(\eta_s^3). \nonumber
\end{align}
The effect of each Hamiltonian is easily understood in the Fock basis of the motional states. 
The single-quantum saturation associated with the qubit degree of freedom plays the fundamental role of allowing the coherent manipulation of single quanta in the motional modes.

The zeroth-order term is the carrier transition Hamiltonian, 
\begin{equation}
\hat H^{(0)} = {\textstyle\frac12}\hbar\Omega'\left(\e^{-i\delta t}\hat\sigma_+ + \e^{i\delta t}\hat\sigma_-\right),
\label{eq:6}
\end{equation}
resonant for $\delta = 0$. The Rabi frequency is modified due to the motional coupling as $\Omega'= (1-\eta_a^2-\eta_b^2)\Omega$. This Hamiltonian induces qubit transitions without affecting the motional state of the ion. It may be used to prepare the qubit quantum state in order to apply suitable control over the motional modes.

The first-order terms involve the blue- and red-sideband transitions of each vibrational mode, through the Hamiltonians
\begin{align}
\label{eq:redblusidebands}
\hat H_{s}^{(1)} = & \;{\textstyle\frac12}\hbar\Omega \left(\e^{-i(\delta +\omega_s) t}\hat\sigma_+\hat s + \e^{i(\delta +\omega_s) t}\hat\sigma_-\hat s^\dag\right)\\
& +{\textstyle\frac12}\hbar\Omega \left(\e^{-i(\delta -\omega_s) t}\hat\sigma_+\hat s^\dag + \e^{i(\delta -\omega_s) t}\hat\sigma_-\hat s\right)\nonumber,	
\end{align}
The first two terms are resonant for $\delta = -\omega_s$ and excite the qubit while annihilating a phonon in mode $\hat s$, and vice-versa; the remaining terms, resonant for $\delta = \omega_s$, promote the excitation of the qubit while creating one additional phonon in the motional mode, and conversely. The CZ paradigm utilizes this Hamiltonian to map the qubit state into one motional mode or to realize conditional logic between them, employing the vibrational mode as an effective two-level ancilla system. 

Our interest here lies mostly in the second-order terms of $\hat H_I$. They entail the creation or annihilation of two phonons at a time together with the excitation or deexcitation of the qubit. The single-mode Hamiltonians are
\begin{align}
\label{eq:tworedblusidebands}
\hat H_{s}^{(2)} = & \;{\textstyle\frac12}\hbar\Omega \left(\e^{-i\delta t}\hat\sigma_+ + \e^{i\delta t}\hat\sigma_-\right)\hat s^\dag\hat s  \\
& + {\textstyle\frac12}\hbar\Omega \left(\e^{-i(\delta +2\omega_s) t}\hat\sigma_+\hat s^2 + \e^{i(\delta +2\omega_s) t}\hat\sigma_-\hat s^\dag{}^2\right)\nonumber\\
& +{\textstyle\frac12}\hbar\Omega \left(\e^{-i(\delta -2\omega_s) t} \hat\sigma_+ \hat s^\dag{}^2 + \e^{i(\delta -2\omega_s) t}\hat\sigma_-\hat s^2\right)\nonumber.
\end{align}
The cross Hamiltonian creates or annihilates pairs of phonons, one in each mode, via the interaction
\begin{align}
& \hat H_{ab}^{(2)} = \frac{\hbar\Omega}{2} \left(\e^{-i(\delta +\omega_a - \omega_b) t}\hat\sigma_+\hat a\hat b^\dag + \e^{i(\delta +\omega_a - \omega_b) t}\hat\sigma_-\hat a^\dag\hat b\right)\nonumber\\
& + \frac{\hbar\Omega}{2}\left(\e^{-i(\delta -\omega_a + \omega_b) t}\hat\sigma_+\hat a^\dag\hat b + \e^{i(\delta -\omega_a + \omega_b) t}\hat\sigma_-\hat a\hat b^\dag\right)\nonumber\\
& +\frac{\hbar\Omega}{2}\left(\e^{-i(\delta +\omega_a + \omega_b) t} \hat\sigma_+ \hat a\hat b + \e^{i(\delta +\omega_a + \omega_b) t}\hat\sigma_-\hat a^\dag\hat b^\dag\right) \nonumber\\
& +\frac{\hbar\Omega}{2}\left(\e^{-i(\delta -\omega_a - \omega_b) t} \hat\sigma_+ \hat a^\dag\hat b^\dag + \e^{i(\delta -\omega_a - \omega_b) t}\hat\sigma_-\hat a\hat b\right).
\label{eq:crosshamiltonian}
\end{align}
Analogously, the $n$\textsuperscript{th}-order term of $\hat H_I$, if considered, would coherently distribute $n$ phonons between the two modes, although with ever decreasing coupling strength of order $\eta_s^n$, which would imply on longer operation times.

\section{Bichromatic CV quantum gates}
\label{sec:bichromatic}

Due to its property of saturating with the absorption of a single quantum, the qubit provides a convenient way to add or subtract individual phonons in motional modes. However, it also prevents the motional state from attaining a fast increase in the number of excitations.

Quantum gates acting on the CV system must be allowed to visit ever higher excitation numbers without saturating, a feature that may require the qubit state to factor out of the interaction. Hence to allow interesting combinations of terms of $\hat H_s^{(1)}$, $\hat H_s^{(2)}$, or $\hat H_{ab}^{(2)}$ to be simultaneously resonant, we consider bichromatic light sources to drive the quantum dynamics~\cite{ms_prl99,sigmaz_00,ms_pra00,roos_njp08}. 

An alternative and more intuitive picture of CV operations in this case makes use of the phase space of position and momentum observables, describing quantum gates in terms of the transformations they produce in the Wigner function of the quantum system. We employ this approach below to describe the quantum operations of the toolbox.

\subsection{Single-mode Gaussian operations}

The single-mode Gaussian operations comprise displacements and squeezers. 

Displacements produce a translation of the phase space, rigidly moving the Wigner function. If the arguments of the single-mode Wigner function are $x_s$ and $p_s$, a displacement with parameter $\alpha = x_\alpha + i p_\alpha$ produces the phase space transformations $x_s\rightarrow x_s - x_\alpha$ and $p_s\rightarrow p_s - p_\alpha$.

Squeezers change the scaling of phase space, compressing and stretching different directions, while respecting the conservation of areas. A squeezer with parameter $\xi = r\exp(2i\theta)$ affects the orthogonal axis $x_\theta = \cos\theta x_s + \sin\theta p_s$
and $p_\theta = -\sin\theta x_s + \cos\theta p_s$ in phase space according to the transformations $x_\theta\rightarrow e^{-r}x_\theta$ and $p_\theta\rightarrow e^r p_\theta$. In the  description in terms of phonons, the most prominent characteristic of a squeezed state is the primacy of even numbers of quanta. 

Although formally implicit in the squeezing operation, we may consider the Fourier transform operation as a distinct quantum gate. In phase space, the Fourier transform gate produces a rotation. It transforms position and momentum according to $x_s\rightarrow x_\theta$ and $p_s\rightarrow p_\theta$ defined above. In the phonon picture, the Fourier gate introduces a number-dependent phase shift in the form $\ket{n_s} \rightarrow e^{in\theta}\ket{n_s}$, where $n_s$ is the phonon number in mode $\hat s$, generating dynamics akin to the free evolution of the oscillator.

In the case of the ion trap processor, a single-mode displacement of mode $\hat s$ is performed by the Hamiltonian $\hat H_s^{(1)}$, given the conditions below.
Two coherent radiation sources with the same intensity and opposite detunings $\delta_1 = -\delta_2 = \delta:=\omega_s$, according to Eq.~(\ref{eq:redblusidebands}), generate the Hamiltonian
\begin{align}
\label{eq:singlemodedisplacement}
\hat H_{s}^{(1)} = & \;i\hbar \hat\sigma_{\phi-\pi/2}\left(\tilde\alpha\hat s^\dag - \tilde\alpha^*\hat s \right),
\end{align}
where $\hat\sigma_{\phi}=\hat\sigma_x\cos\phi+\hat\sigma_y\sin\phi$ and the relative phase between the two frequency components of light is $2\phi$. This phase controls the displacement parameter per unit time, given by $\tilde\alpha = \Omega\e^{i\left(\phi-\frac\pi2\right)}$. 
The realization of a single-mode displacement operation requires the qubit quantum state to be initialized in an eigenstate of the operator $\hat\sigma_{\phi-\pi/2}$, it can be done by a resonant pulse as in Eq.~(\ref{eq:6}). Let us adopt the convention that the qubit eigenstate $\ket{+}_{\phi-\pi/2}$ with positive eigenvalue is chosen. Then the displacement $\hat D(\alpha):=\exp\left(\alpha\hat s^\dag - \alpha^*\hat s\right)$ of mode $\hat s$ by the amplitude $\alpha = \eta_s^2\tilde\alpha t$ is produced by the evolution operator
$\hat D_s  = \exp\left( -i\eta_s^2\hat H_{s}^{(1)}t/\hbar \right)$.

The squeezing operation is realized by considering the Hamiltonian $\hat H_{s}^{(2)}$. The bichromatic field with detunings $\delta_1 = -\delta_2 = \delta:= 2\omega_s$ will produce according to Eq.~(\ref{eq:tworedblusidebands}) the non-linear dynamics
\begin{align}
\label{eq:singlemodesqueezer}
\hat H_{s}^{(2)} = & \;i\hbar \hat\sigma_\phi\left(\tilde\xi^*\hat s^2 - \tilde\xi\hat s^\dag{}^2 \right),
\end{align}
where $\tilde\xi = \Omega\e^{i(\phi-\pi/2)}$ is the squeezing parameter per unit time. Provided the qubit is prepared in the eigenstate $\ket{+}_{\phi}$, acting with this Hamiltonian on the motional state for time $t$ will realize the single-mode squeezing operator $S(\xi) := \exp\left(\xi^*\hat s^2 - \xi\hat s^\dag{}^2\right)$, with squeezing parameter $\xi = \eta_s^2\tilde\xi t$. The corresponding evolution operator is $\hat S_s = \exp\left(-i\eta_s^2\hat H_{s}^{(2)}t/\hbar\right)$.
Although not systematically investigated, quantum noise reduction has been observed in the motional state of trapped ions~\cite{wineland_prl96,squeezingHome_nature15}, revealing the potential of the ion trap to produce large amounts of squeezing for quantum computing.


The Fourier transform gate is realized by the frequency setting $\delta=0$, i.e. by a monochromatic laser tuned to the qubit transition. The Hamiltonian reads as
\begin{equation}
\hat H_{s}'^{(2)} = {\textstyle\frac12}\hbar\Omega \hat\sigma_\phi\hat s^\dag\hat s,
\end{equation}
and produces the evolution $\hat F = e^{i\theta\hat s^\dag\hat s}$, where the rotation phase is $\theta = \eta_s^2\Omega t/2$, provided the qubit remains in the state $\ket{+}_\phi$. 

\subsection{Two-mode Gaussian operations}

Gaussian operations acting on two vibrational modes produce transformations in linear combinations of modal operators. They can be either passive, promoting the exchange of quanta between modes, or active, in which case quanta are concomitantly added or removed from both modes in a correlated way. 

The passive operation is the beam splitter (named after its optical counterpart), a quantum gate that coherently combines the modal operators by amounts that vary with the interaction time. The active operation of a two-mode squeezer produces correlated pairs of quanta, in the phonon picture, or EPR-like entangled states in the phase space picture~\cite{epr_pr35}.

Two-mode operations are realized in the ion trap by the cross Hamiltonian $H_{ab}^{(2)}$. A bichromatic field may modify Eq.~(\ref{eq:crosshamiltonian}) to produce two types of dynamics.
For the first dynamics, we choose radiation frequencies such that $\delta_1 = -\delta_2 := \delta = \omega_a-\omega_b$ (we assume $\omega_a>\omega_b$ for definiteness). The resulting Hamiltonian reads as
\begin{align}
& \hat H_{ab}^{(2)} = \hbar\Omega \,\hat\sigma_\phi\left(\e^{-i\phi}\hat a\hat b^\dag + \e^{i\phi}\hat a^\dag\hat b\right).
\label{eq:beamsplitter}
\end{align}
The beam splitter transformation is realized by the evolution operator $\hat I_{ab} = \exp\left(-i\eta_a\eta_b\hat H_{ab}^{(2)} t/\hbar\right)$. It promotes the interference of modal operators in the form $\hat a\rightarrow \hat a\cos(\phi t) + \hat b \sin(\phi t)$ and $\hat b\rightarrow -\hat a\sin(\phi t) + \hat b \cos(\phi t)$, where $\phi= 2\eta_a\eta_b\Omega$. For instance, modes interfere maximally at half this time, for  $t = \pi/(2\phi)$; for double that amount of time, the quantum state of one mode is mapped into the other, and vice-versa, coherently exchanging their local quantum states. In case the two-mode quantum state is initially separable, the beam splitter dynamics will entangle the modes unless the initial state is a coherent state.

The second type of two-mode dynamics involves the choice of detunings $\delta_1 = -\delta_2 := \delta = \omega_a+\omega_b$. The resonant terms of Eq.~(\ref{eq:crosshamiltonian}) produce the interaction Hamiltonian
\begin{align}
\hat H_{ab}^{'(2)} = \hbar\Omega \,\hat\sigma_\phi\left(\e^{-i\phi}\hat a\hat b + \e^{i\phi}\hat a^\dag\hat b^\dag\right).
\label{eq:twomodesqueezer}
\end{align}
The two-mode squeezer $\hat E_{ab} = \exp\left(i\hat H_{ab}^{'(2)} t/\hbar\right)$ produces EPR-type entangled states between the motional modes when acting over the oscillator ground state. As in previous cases, the relative phase between the laser frequency components controls the operation phase as long as the qubit degree of freedom is prepared in the eigenstate $\ket{+}_{\phi}$.

\subsection{Non-Gaussian single- and two-mode operations}

Gaussian operations are restricted to produce Gaussian quantum states when starting from one. To produce more general quantum states and achieve computational universality, at least one non-Gaussian operation is needed~\cite{bartlett_prl02,bartlett_pra02}. In fact, the ion trap quantum processor provides a wide variety of non-Gaussian operations, made available by the strong non-linearity of the qubit degree of freedom, which saturates with a sole quantum. The number of phonons in the CV degrees of freedom can thus be increased by discrete amounts by coupling it with the qubit, thus producing non-Gaussian quantum states.

The most convenient operation of single-mode phonon creation or annihilation stems from the red- and blue-sideband interactions produced by $\hat H_s^{(1)}$ [Eq.~(\ref{eq:redblusidebands})]. They can be realized with a monochromatic laser beam with frequency $\delta = \omega_s$ (blue sideband) or $\delta = -\omega_s$ (red sideband), respectively, yielding
\begin{align}
\hat H_{s,\text{blue}} & = {\textstyle\frac12}\hbar\Omega \left(\hat\sigma_+\hat s^\dag + \hat\sigma_-\hat s\right),	\\
\hat H_{s,\text{red}} & = {\textstyle\frac12}\hbar\Omega \left(\hat\sigma_+\hat s + \hat\sigma_-\hat s^\dag\right).
\end{align}
We denote the respective evolution operators as $\hat B_s = \exp\left(-i\eta_s\hat H_{s,\text{blue}}t/\hbar\right)$ and $\hat R_s = \exp\left(i\eta_s\hat H_{s,\text{red}}t/\hbar\right)$.
Both interactions show linear coupling strength on $\eta_s$, making them stronger in comparison with the Gaussian operations considered previously. This property favors the creation or annihilation of any number of phonons in a stepwise process, by applying blue or red sidebands intercalated by $\pi$-pulses in the carrier transition ($\delta=0$). In this manner, the Fock state $\ket{n}$ can be generated.

In the case of two modes, the cross Hamiltonian $\hat H_{ab}^{(2)}$ allows the simultaneous creation or annihilation of two phonons, one in each mode, by also employing a laser of single frequency. In the first case, a laser with detuning $\delta = \pm(\omega_a -\omega_b)$ will remove one phonon from one mode and create one phonon in the other, promoting the coherent exchange of a single excitation between modes. 

To avoid entanglement between qubit and vibrational modes, creation or annihilation of phonons must be realized over suitable initial states, such as the ground state of the oscillator. In this manner, non-Gaussian features can be used as resources introduced at certain steps of the computation obeying such constraint (e.g. at its beginning).

Finally, a third vibrational mode can be employed as an ancilla to perform operations on the other modes. For instance, in the case of Gaussian operations, a displacement operation can be realized by writing a coherent state with large amplitude on the ancilla mode and applying the beam splitter operation for a short duration to coherently combine its quantum state with that of mode $\hat a$ or $\hat b$. Non-Gaussian features could also be written in the third mode and then introduced in other modes by the two-mode Gaussian operators mentioned previously. Operations such as sum or subtraction of phonons could be realized in a similar manner.

\subsection{Two-mode conditional operations}

Controlled operations involving two modes are intended to change the quantum state of one mode conditioned on the state of another, usually generating entanglement~\cite{reviewbraunsteinvanloock_rmp05}. One example of such operation is the controlled displacement, represented by the operator
\begin{equation}
\hat C_{\hat x_a}=\exp(-i\hat x_a\otimes\hat p_b)\;,
\label{eq:controlledgate}
\end{equation}
where the order of the systems is {\it control} $\otimes$ {\it target}. For instance, if both modes start as independent coherent states, this gate displaces the average momentum of the target mode by the average position of the control mode. Other types of controlled gates can be devised by taking Eq.~(\ref{eq:controlledgate}) as model.

In the ion trap, controlled gates can be deployed by two bichromatic lasers, i.e. by tetrachromatic light. In fact, by the form of Eq.~(\ref{eq:controlledgate}), conditioned dynamics can be seen to involve a combination of beam splitter and squeezing operations. A tetrachromatic laser with frequency components $\delta_1=-\delta_2:=\omega_a-\omega_b $ and $\delta_3=-\delta_4:=\omega_a+\omega_b$ with the same intensity  gives rise to the dynamics
\begin{equation}
\hat H_{ab}^{''(2)} = \hbar\Omega \,\hat\sigma_\phi\hat x_a\hat x_{\phi,b},
\end{equation}
where $\hat x_{\phi,b} = e^{-i\phi}\hat b+e^{i\phi}\hat b^\dag = \hat x_b\cos\phi + \hat p_b\sin\phi$ is a rotated coordinate in the phase space of mode $\hat b$. The evolution operator is $\hat C_{ab} = \exp\left(-i\eta_a\eta_b\hat H_{ab}^{''(2)}t/\hbar\right)$.

\subsection{Schwinger map}

The algebra of angular momentum can be realized in terms of operators associated with two harmonic oscillators, according to the map $\hat J_+ = \hat a^\dag\hat b$ and $\hat J_- = \hat a\hat b^\dag$, where $\hat J_\pm = \hat J_x \pm i\hat J_y$ are the raising and lowering operators defined in terms of the $x$- and $y$-components of the angular momentum vector. The corresponding $z$-component is represented by the operator $\hat J_z = (\hat a^\dag \hat a - \hat b^\dag \hat b)/2$. Each eigenvalue of the angular momentum magnitude $\hat J^2$ corresponds to a fixed total number $N = n_a + n_b$ of phonons available in the two modes. Eigenstates of $\hat J^2$ can be represented by an angular momentum vector over a sphere. The poles of the sphere are the eigenstates $\ket{N}$ and $\ket{-N}$ of $\hat J_z$. They correspond to the harmonic oscillator states $\ket{N}_a\ket{0}_b$ and $\ket{0}_a\ket{N}_b$, respectively. The other eigenstates of $\hat J_z$ exist as superpositions of Fock states $\ket{n}_a\ket{N-n}_b$, where $n$ is the difference of quanta between modes $\hat a$ and $\hat b$ (equal to half the eigenvalue of $\hat J_z$).

It might be useful sometimes to think of the two-mode quantum operations described above as the manipulation of such angular momentum vector, even though in most cases the two-mode motional quantum state will not be in a superposition of Fock states satisfying the property $n_a+n_b=N$. The Schwinger map is particularly suitable to treat the beam splitter operation, since it conserves the total number of excitations in the two modes.
In the Schwinger map, the beam splitter Hamiltonian of Eq.~(\ref{eq:beamsplitter}) reads as
\begin{equation}
\hat H_{ab}^{(2)} = \hbar\Omega \,\hat\sigma_\phi\hat J_\phi,
\label{eq:beamsplitterschwinger}
\end{equation}
where $\hat J_\phi = \cos\phi\hat J_x + \sin\phi\hat J_y$.
Hence the beam splitter operation can be interpreted as a rotation of the angular momentum vector existing in the sphere associated with an eigenstate of $\hat J^2 = \hat J_x^2 + \hat J_y^2+\hat J_z^2$. For quantum states inhabiting more than one sphere, the beam splitter realizes a coherent superposition of rotations, one in each sphere.

\section{Ion trap CVQC toolbox}
\label{sec:toolbox}

\subsection{Architecture}

The realization of CVQC with the motional modes of trapped ions requires the ability to select specific quantum dynamics by tuning the properties of the manipulation laser. As presented above, quantum gates can be implemented with monochromatic radiation or using a combination of frequency components commensurate with the frequencies of vibration. They are selected by radiation frequency (detuning) and direction of propagation (Lamb-Dicke parameter).

\begin{figure}[htbp]
\centerline{\includegraphics[width=0.95\columnwidth]{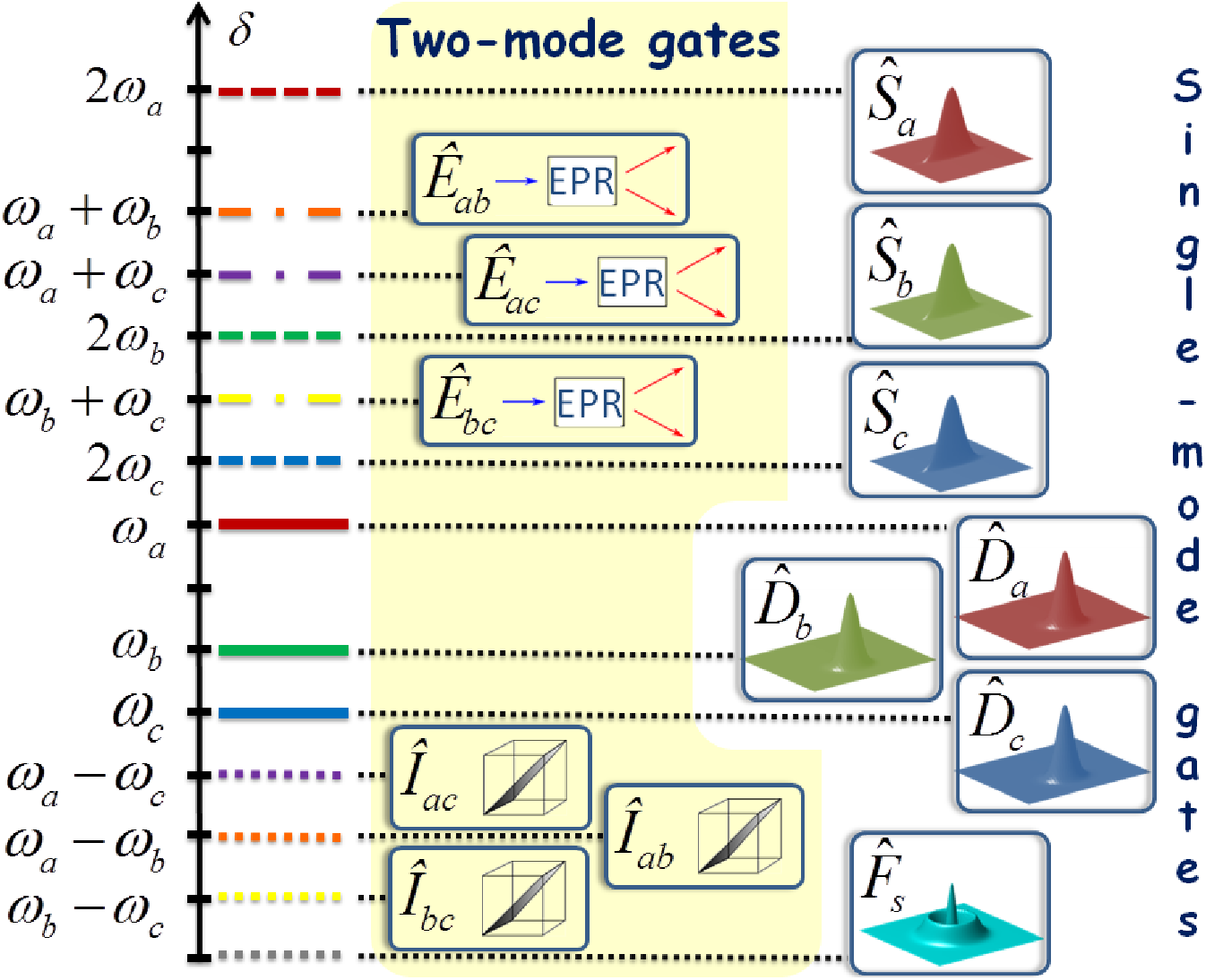}}
\caption{Toolbox of Gaussian operations available for the manipulation of the vibrational state of a single trapped ion. The desired quantum gate is selected by radiation frequency. With the exception of the Fourier transform operation $\hat F_s$, all quantum gates require bichromatic radiation with $\delta_1=-\delta_2=\delta$. Possible values of detuning correspond to any of the: vibrational frequencies $\omega_s$ (continuous lines), double those frequencies (dashed lines), subtraction (dotted lines) or sum (dash-dot lines) of pairs of frequencies. The corresponding Gaussian operations are displacements $\hat D_s$, squeezers $\hat S_s$, beam splitters $\hat I_{ss'}$, and two-mode squeezers $\hat E_{ss'}$, respectively, where $s,s'\in\{a,b,c\}$. In choosing the ratio of vibrational frequencies, we have adopted the proportion $\omega_a:\omega_b:\omega_c = 7:5:4$. }
\label{fig:toobox}
\end{figure}

The quantum gates considered in Sec.~\ref{sec:bichromatic} form the basic toolbox to perform the coherent manipulation of the vibrational modes. Fig.~\ref{fig:toobox} shows the Gaussian operations together with the bichromatic laser detunings required for their realization in the case that all three vibrational modes of a single trapped ion are employed (the label $s$ now reads $s\in\{a,b,c\}$). 

The single-mode quantum gates comprise the displacement $\hat D_s$, squeezer $\hat S_s$, Fourier transform $\hat F_s$, blue- $\hat B_s$, and red-sideband $\hat R_s$ operations. Their necessary laser frequency components and evolution operators are summarized as:
\begin{align}
\begin{array}{ll}
\delta_1=-\delta_2 = \omega_s: & \; \hat D_s = e^{-\frac{i}{\hbar}\eta_s^2\hat H_{s}^{(1)}t},\\
\delta_1=-\delta_2 = 2\omega_s: & \; \hat S_s = e^{-\frac{i}{\hbar}\eta_s^2\hat H_{s}^{(2)}t}, \\
\delta = 0: & \; \hat F_s = e^{-\frac{i}{\hbar}\eta_s^2\hat H_{s}'^{(2)}t}, \\
\delta = \omega_s: & \; \hat B_s = e^{-\frac{i}{\hbar}\eta_s\hat H_{s,\text{blue}}t}, \\
\delta = -\omega_s: & \; \hat R_s = e^{-\frac{i}{\hbar}\eta_s\hat H_{s,\text{red}}t}. \\
\end{array}
\end{align}

Two-mode operations must be applied to pairs of modes. To account for the three combinations of mode pairs, we introduce a second index $s'\in\{a,b,c\}$. The two-mode gates are the beam splitter $\hat I_{ss'}$, the two-mode squeezer $\hat E_{ss'}$, and the general conditional operation $\hat C_{ss'}$. They require the radiation frequencies:
\begin{align}
\begin{array}{ll}
\delta_1=-\delta_2 = \omega_s - \omega_{s'}: & \hat I_{ss'} = e^{-\frac{i}{\hbar}\eta_s\eta_{s'}\hat H_{ab}^{(2)} t},\\
\delta_1=-\delta_2 = \omega_s + \omega_{s'}: & \hat E_{ss'} = e^{-\frac{i}{\hbar}\eta_s\eta_{s'}\hat H_{ab}^{'(2)} t}, \\
\left\{
\begin{array}{l}
\delta_1=-\delta_2 = \omega_s - \omega_{s'} \\
\delta_3=-\delta_4 = \omega_s + \omega_{s'}
\end{array}
\right.:
 & \hat C_{ss'} = e^{-\frac{i}{\hbar}\eta_s\eta_{s'}\hat H_{ab}^{''(2)}t}. \\
\end{array}
\end{align}
Three types of gates utilize monochromatic radiation (Fourier transform $\hat F_s$, blue-sideband $\hat R_s$ and red-sideband $\hat R_s$), four require bichromatic lasers (displacement $\hat D_s$, squeezer $\hat S_s$, beam splitter $\hat I_{ss'}$, and two-mode squeezer $\hat E_{ss'}$), and one employs tetrachromatic light (the conditional operation $\hat C_{ss'}$).

Different operations are selected by laser detuning $\delta$ and propagation direction (through $\eta_s$), so as to keep off-resonant terms of the complete Hamiltonian sufficiently detuned in order to avoid exciting population in undesirable quantum states. Hence for a given pair of modes the proposed CVQC architecture requires the frequencies $\omega_s$, $\omega_{s'}$, $2\omega_s$, $2\omega_{s'}$,  $\omega_s-\omega_{s'}$, and $\omega_s+\omega_{s'}$ to be incommensurate and sufficiently separated. For a processor based on the three motional modes of a single trapped ion, a total of 12 incommensurate frequencies must be available. For instance, the oscillation frequencies $\omega_a = 7$~MHz, $\omega_b = 5$~MHz, and $\omega_c = 4$~MHz (i.e. in the proportion 7:5:4 adopted in Fig.~\ref{fig:toobox}) would furnish 1~MHz as the free spectral interval between quantum gates. These  vibrational frequencies can be achieved by engineering the trap potential through the electrodes geometry and the magnitudes of the applied external voltages. Asymmetric trap designs producing vibrational frequencies as high as 50~MHz have been demonstrated~\cite{winelandtrapasym_pra95}.

Control of the Lamb-Dicke parameters $\eta_s$, although not necessary for all quantum gates, also helps avert off-resonant excitations. Since these parameters depend on the direction of laser propagation, it is possible to mitigate undesirable excitations by employing laser beams propagating in suitable directions. The Fourier transform operation, however, requires the ability to control the Lamb-Dicke parameters by laser direction, given that it utilizes the qubit carrier frequency for the manipulation of all modes. The Fourier transform can also be made to act simultaneously on more than one mode by adapting the same idea.

The CVQC ion trap architecture utilizes the qubit degree of freedom as mediator for the interaction between modes, and as such it must remain separable from the vibrational modes once CV gates are applied. That requirement is automatically fulfilled by all Gaussian quantum gates, since in those cases the qubit remains in an eigenstate of a Pauli operator throughout the quantum evolution. The only cases in which the fulfillment of this requirement must be verified are those involving the application of non-Gaussian operations (i.e. blue- and red-sidebands), for which the saturation properties of the qubit are harnessed and thus play a role in entangling it with the vibrational modes.

Considering a single motional mode for the sake of the argument, blue- and red-sideband operations induce transitions between the basis states $\ket{g,n_s}\leftrightarrow\ket{e,n_s\pm1}$ with effective coupling rates varying as $\Omega_{n_s}=\Omega_0\sqrt{n+1}$, where $\Omega_0=\eta_s\Omega$ is the Rabi frequency for the fundamental transition $\ket{g,0}\leftrightarrow\ket{e,1}$, in case of the blue-sideband operation, or $\ket{e,0}\leftrightarrow\ket{g,1}$ for the red-sideband. For instance, the creation of an additional phonon on the quantum state $\ket{g,n_s}$ would require a $\pi$ pulse, corresponding to the interaction time $\tau_{n_s}=\pi/\Omega_0\sqrt{n_s+1}$ which depends on $n_s$.
Thus a general single-mode vibrational quantum state initially separable from the qubit state and written as $\ket{\psi_s}=\sum_{n_s}d_{n_s}\ket{n_s}$ would become entangled with the qubit after application of any of these non-Gaussian pulses.

To avoid this situation, non-Gaussian operations cannot be applied to any quantum state, but must be restricted to initial states capable of satisfying the requirement of factorization of the qubit upon completion of the quantum gate. The ground state is an obvious choice of initialization. A sequence of blue-sideband pulses and carrier pulses can then be applied to produce any Fock state $\ket{n_a,n_b,n_c}$.

In the first order Lamb-Dicke approximation ($\eta\ll1$), the phonon number of other vibrational modes different than the ground state, does not affect the dynamics of the system. But, as a second  order effect, the occupation of these  modes changes the ion-light coupling strength, which causes an effective  fluctuation on Rabi frequency as the number of phonons follows a thermal distribution after cooling \cite{Wineland:1997mg,Poschinger2012}. However, the Lamb-Dicke parameter, which depends on the angle between the direction of oscillation of the ion and the propagation of the laser, can and should be used to minimize undesired excitations of other modes of vibration.

In truth, the class of non-Gaussian quantum states available can be increased by employing the Schwinger map. Considering a pair of modes, states with the form $\ket{n_s,0_{s'}}$ possess well defined value of $\hat J_z$. The beam splitter operation can then be applied to distribute the $n_s$ quanta between the modes while keeping the qubit separable. 
Then two-mode non-Gaussian states are also available which are built as superpositions of the basis states $\ket{n_s,n_{s'}}$ satisfying $n_s+n_{s'}=N'$, where $N'$ is a constant. 


\subsection{Dimension of the motional Hilbert space}

It is a useful exercise to estimate the potential limits of CVQC with the motional modes of a trapped ion. The number of modes would probably be limited to a few, owing to the requirement of frequency selectivity in the coherent manipulation. Here we estimate the limitations of the most basic architecture composed of a single trapped ion and three motional modes.

Let us first estimate the expected performance of the ion trap motional modes with proven technology. Typical large ion traps, with electrode distances on the order of 1~mm, although not built for the purpose of CVQC, can currently handle at least $N\approx 100$~phonons according to experiment~\cite{quantumwalkions_10}. The Hilbert space readily available in the three motional modes of such traps would have dimension $D\approx 10^6$. The equivalent number $N_\mathrm{qubit}$ of ions needed to produce a Hilbert space of the same dimension in the CZ paradigm would be $N_\mathrm{qubit}\approx \log 10^6/\log 2 \approx 20$. Hence we can safely estimate that transitioning  to the CVQC could in principle increase the Hilbert space available in the current ion trap processor to values much larger than the best limits demonstrated to date~\cite{qubits14_prl11}.

Ultimately, the size of the Hilbert space associated with each mode is limited by anharmonicities in the trap potencial, which make the energy separation between adjacent Fock states dependent on the number of phonons. Since actual implementations of the harmonic potential can only be valid in a restricted physical volume, anarmonicities will introduce a maximum phonon number cap $N$ per mode. Hence for a single trapped ion the three vibrational modes available would entail a Hilbert space of dimension $D\approx N^3$.
To estimate practical values of $N$, we may consider the typical length of a long ion chain available with current technology ($\ell\approx 100$~$\mu$m) as providing the maximum attainable length for the wavefunction of a single trapped ion. Imposing $\ell\approx\sqrt{N+1}x_s$, the phonon cap would thus be limited to $N\approx(\ell/x_s)^2\approx 10^8$. The three modes would together visit a Hilbert space of dimension $N\approx 10^{24}$. The same Hilbert space could be accessed in the CZ paradigm by employing $N_\mathrm{qubit}\approx 80$ ionic qubits. This value represents the limitations imposed by the hardware on our CVQC architecture in the simplest scenario using proven technology (i.e. wherein a single trapped ion is utilized and the harmonic potential is not optimized to cover a larger volume in space).

A more stringent limit would consider the onset of undesirable effects on the CV quantum gates as they start to show dependence on the phonon number for large $N$. In fact, higher order terms can be neglected in the expansion of Eq.~(\ref{eq:Hexpansion}) only if $\eta_s^2N\ll1$. We hence consider $N\approx 0.01/(\eta_s^2)$ as the reasonable phonon cap of our architecture. For laser excitation of fixed frequency, the Lamb-Dicke parameter scales with the oscillator frequency as $\omega_s^{-1/2}$. Increasing the trap stiffness thus allows for a larger Hilbert space while decreasing the interaction strength and hence extending the duration of quantum gates, which limits the maximum number of gates whithin the coherence time of the motional modes. This latter side effect could be compensated by increasing the laser power as long as off-resonant excitations can be neglected. A credible scenario considering the balance between gate speed and trap stiffness would put the Lamb-Dicke parameter at values around $\eta_s\approx10^{-3}$, allowing for the three-mode Hilbert space to achieve the realistic dimension of $D\approx10^{12}$. 
The equivalent number of qubits would be in this case $N_\mathrm{qubit}\approx 40$. Interesting applications in quantum simulations already exist for a configuration space of this size~\cite{qbitenconding100photon_science13,cverrorcorr_prl98,lloyd_prl99,gottesmanencoding_pra01}.

Another important aspect to consider in this trap architecture is the heating effect on the coherence time. The Innsbruck group has seen a coherence time for the vibrational state of the order of $100$ ms \cite{Schmidt-Kaler2003} in the center of mass mode, which is consistent with voltage fluctuations expected to be around $10^{-5}$, they measure the increase in the number of phonons by the observation of the Rabi oscillation signal on the blue sideband in a Ramsey experiment. Specifically, it can be seen that the creation time of one phonon is measured again by the Innsbruck group, being about $140$ ms, and the decoherence time of the information stored in the vibrational mode of the order of $85$ ms \cite{Monz2009}. Furthermore, the Oxford group observed in a single calcium ion a coherence time of $182$ ms for the motion state, limited by movement heating of around $3$ quanta/s \cite{Lucas2007}. However, the Innsbruck group found that in the case of a chain of ions, for axial breathing mode (and other higher axial modes) are more typical the coherence times of about $5$ ms \cite{Roos2008}, therefore is more clear the advantage of using a single ion to protect the states of motion from the decoherence effect.

The problem of spectral crowding will not be, in principle, present in our system since our proposal is indeed made for a single ion, although we do not disregard as a perspective to explore the possibility of extending it to more than one atom. However if we consider the ability to scale the system for more than one ion, a possible way to overcome the problem of spectral crowding is adopting a modular design, i.e. the decomposition of the device in a larger number of interconnected traps, either using a complicated geometry to allow ions exchange from a trap to another or following a hybrid system, for example making use of photonics interfaces which according to recent studies are proving to be less and less disadvantageous in terms of noise and speed efficiency of operations, compared to an isolated ion crystal \cite{Li2016}.

As a side advantage, the vibrational modes of trapped ions can be made to attain a large Hilbert space while producing modest increase in the \textit{physical volume}  occupied by the quantum system (and thus decreasing the experimental complexity), a feature that should help protect it from environmental decoherence. 

\subsection{Measurement of the motional quantum state}

It is a daunting challenge to completely characterize the quantum state of large Hilbert spaces, a hurdle inherent to any architecture of a quantum computer. But it is often not necessary. A good quantum algorithm must yield as the result of computation an answer that does not require complete quantum state reconstruction~\cite{nmrQuantumAlgorithm_98,jozsaBlatt_nature02,fourierWineland_science05}.

In the case of the ionic vibrational modes, the information directly available to measurement is either \textit{(i)} the populations of Fock states or \textit{(ii)} values of the Wigner function at any point in phase space. The possibility to choose between two different sets of quantum observables brings an additional flexibility to this physical system.
Following \textit{(i)}, if we write the motional quantum state in the Fock basis of Eq.~(\ref{eq:quantumstatefockbasis}), then the populations $|d_{n_an_bn_c}|^2$ are available to measurement; In case the phase space picture is better suited to the kind of computation being performed, scenario \textit{(ii)} puts the value of the Wigner function at any point $(x_a,p_a,x_b,p_b,x_c,p_c)$ as determined by the parity of the phonon distribution, a quantity that can be obtained without the need to recover the full population of Fock states, but by directly employing the qubit degree of freedom~\cite{lutterbachdavidovich_prl97}. Direct measurements of the Wigner function can also be utilized to infer general properties of the quantum state~\cite{paris_pra31,cavesralphbosonsampling2016}. We develop more on each possibility in the following.

As noted previously, in scenario \textit{(i)} the Fourier transform gate acting on a single mode presents phonon-number dependent coupling in transitions of the type $\ket{g,n_s}\leftrightarrow\ket{e,n_s\pm1}$. This means that driving the qubit with a resonant laser will produce Rabi oscillation composed of different harmonic components. The Rabi frequency associated with the general transition $\ket{g,n_a,n_b,n_c}\leftrightarrow\ket{e,n_a,n_b,n_c}$ is
\begin{equation}
\Omega_{n_an_bn_c} = \Omega' - \Delta\Omega_{n_an_bn_c},
\label{eq:rabifrequencycarrier}
\end{equation}
showing a fast component $\Omega'$ (employed for short times to manipulate the qubit in the Bloch sphere during the computational stage) and a slowly varying envelope with frequency $\Delta\Omega_{n_an_bn_c} = \Omega_0\sum_s\eta_s^2 n_s$ that contains the phonon number dependence and thus the information on the Fock state populations~\cite{matosfilhovogel_prl96,davidovich_pra96}. While the same is true for driving the blue- or red-sideband transitions, the Fourier gate has the advantage of providing linear dependence on $n_s$, in contrast with the $\sqrt{n_s}$ dependence of the $\hat B_s$ and $\hat R_s$ operations. 

Hence by driving the carrier transition ($\delta=0$) and recording the ion excitation as a function of time, one is able to distinguish among the $N^3$ different populations given that all frequencies $\Delta\Omega_{n_an_bn_c}$ are different, a condition achieved by tuning the Lamb-Dicke parameters $\eta_s$. The actual implementation of these ideas will require defining the maximum time interval $T$ available for driving the qubit Rabi oscillation (limited by qubit decoherence and phonon number population integrity) and choosing slightly different values for $\eta_s$ while respecting the condition $\Omega_0\sum_s\eta_s^2 N\ll\Omega'$ for the phonon cap. For instance, one could impose $\eta_c = \eta_b +\delta\eta$ and $\eta_b=\eta_a+\delta \eta$ while requiring the fastest phonon component $\Delta\Omega_{001}$ to be much smaller than $\Omega'$, e.g. by the conditions $\Delta\Omega_{001} = \Omega'/n_\mathrm{flops}$, where $n_\mathrm{flops}\gg1$ is a constant, and $\Delta\Omega_{100} \approx \Omega'/(n_\mathrm{flops}+\Delta n_\mathrm{flops})$, where the constant $\Delta n_\mathrm{flops}\sim 1$ allows one to distringuish among phonon modes. On the other limit, the value of $T$ fixes the smallest detectable fringe displacement, and thus the phonon cap $N$, by the condition $\Delta\Omega_{NNN}\approx 2\pi/T$. In this manner, the experimental context defines the possible values of $\eta_s$ and $N$. As a numerical example, supposing the experimental values $\Omega_0\approx 2\pi\times$~10~krad/s and $T\approx 100$~ms~\cite{trapinnsbruck_apb03}, one obtains $\eta_s\approx10^{-1}$ and $N\approx100$; The Hilbert space would comprise $10^6$ basis states, imprinting the need to detect the same amount of Fourier frequencies in the envelope of $10^4$ Rabi oscillations of the qubit (hence each oscillation should be sampled by $\approx 10^2$ measurements). Increasing the maximum number of distinguishable phonon number populations requires faster driving capability of the carrier frequency $\Omega_0$~\cite{monroeultrafast_prl10}, or even larger $T$ for the measurement of the oscillation series~\cite{suppressionheatingMonroe_prl06,suppressionheatingChuang_prl08}.

Scenario \textit{(ii)} involves using the property of the Wigner function whereby its value at the origin of phase space is proportional to the average of the parity operator $\hat P = e^{i\pi\sum_s\hat s^\dag\hat s}$ on the vibrational modes \cite{parityWigner}, i.e.
\begin{equation}
W(\{0\}) = \frac{2}{\pi}\sum_{n_a,n_b,n_c} (-1)^{n_a+n_b+n_c} |d_{n_an_bn_c}|^2.
\end{equation}
In this approach, coherences can also be accessed. The Wigner function $W(x_a,p_a,x_b,p_b,x_c,p_c)$ at any other point in phase space can be measured by displacing the quantum state conveniently.

The Fourier gate provides the mapping of the phonon populations into the qubit excitation~\cite{matosfilhovogel_prl96,davidovich_pra96}. In the tridimensional case, a convenient choice of Lamb-Dicke parameters for this type of measurement is $\eta_a=\eta_b=\eta_c:=\eta$. The Rabi frequency of Eq.~(\ref{eq:rabifrequencycarrier}) can be written in this case as
\begin{equation}
\Omega_{n_an_bn_c} = \Omega_0 - (1 + n)\Delta\Omega,
\label{eq:rabifrequencycarrieralternative}
\end{equation}
where the envelope beatnote frequency is $\Delta\Omega = \eta^2\Omega_0$ and $n=n_a+n_b+n_c$. 
Upon application of the Fourier gate, the motional quantum state of Eq.~(\ref{eq:quantumstatefockbasis}) becomes entangled with the qubit, supposed initially in the ground state, yielding
\begin{align}
\ket{\Psi(t)} = \sum_{n_a,n_b,n_c=0}^{N}& d_{n_an_bn_c}\Big(\cos(\Omega_{n_an_bn_c}t)\ket{g}\ket{n_a,n_b,n_c}\nonumber\\
& +\sin(\Omega_{n_an_bn_c}t)\ket{e}\ket{n_a,n_b,n_c}\Big).
\end{align}
Choosing the gate application time $t=t_0$, so that we have $|\cos(\Delta\Omega t)|=1$ for odd $n$ and null otherwise, and commensurate frequencies such that $\Omega_0/\Delta_\Omega = 4m$ ($m$ integer), we find
\begin{align}
\ket{\Psi(t_0)} = \sum_{n}& \Big(d_{2n+1}(-1)^{2n+1}\ket{g}\ket{n_a,n_b,n_c}\nonumber\\
& +d_{2n}(-1)^{2n}\ket{e}\ket{n_a,n_b,n_c}\Big),
\end{align}
where $d_n = d_{n_an_bn_c}$ with the restriction $n_a+n_b+n_c=n$. The parity can be obtained by tracing out the motional modes. The final qubit density matrix presents a sum of the diagonal elements $|d_{n_an_bn_c}|^2$ with $n$ odd in the sector $\ket{g}\bra{g}$ and vice-versa. The Wigner function at the origin of phase space then reads as~\cite{matosfilhovogel_prl96,davidovich_pra96}
\begin{equation}
W(\{0\}) = \frac{2}{\pi}\left(P_e - P_g\right),
\end{equation}
where $P_e = \sum_n|d_{2n}|^2$ and $P_g = \sum_n|d_{2n+1}|^2$ are respectively the populations of $\ket{e}$ and $\ket{g}$ qubit states.

In case one wishes to increase the phonon cut-off $N$, the Rabi frequency associated with the phonon state readout may become untenably high with regular pulses. There is a variety of tools developed by the nuclear magnetic resonance community that can be applied to overcome this limitation
\cite{DANTE1, DANTE2, Chirped1, Chirped2, Chirped3, Tannus1, Tannus2, HS1, HS2}. All these techniques rely on frequency
modulation to increase the pulse bandwidth excitation with limited power. The DANTE experiment consists of short pulses
equally spaced in time and interleaved at different frequencies in the bandwidth to be excited~\cite{DANTE1, DANTE2}.
Recently, a similar method had been proposed in the context of trapped ions to readout the motional state~\cite{Johnson2015}.
The main idea is to perform a Ramsey experiment using spin-dependent kicks (SDK), whose propagator is
$\hat{U}_{SDK} = \hat{\cal{D}}(i\eta)\hat{\sigma}_{+} + \hat{\cal{D}}(-i\eta)\hat{\sigma}_{-}$ and where $\hat{\cal{D}}$ is the displacement
operator of a single mode. These SDK's are concatenated to generate a larger effective SDK, $\hat{U}_{SDK}^{N'}$, where $N'$ is the number of SDK's.
In principle, such method should be able to sense motional states up to $\bar N \approx 10^9$~\cite{Johnson2015}.

Also of particular interest are the so called `adiabatic pulses' \cite{Tannus1, Tannus2}, extensively employed in magnetic
resonance imaging. These pulses can have any frequency and/or amplitude modulation as long as a simple adiabatic condition is fulfilled which relates pulse strength to how fast its frequency modulation occurs. When properly designed, the main
advantages of the adiabatic pulses are: \textit{(i)} the excited bandwidth depends only on the frequency sweep during the
pulse and \textit{(ii)} their insensitivity to inhomogeneities in the strength of the pulses and off-resonance effects for a given pulse
power~\cite{Tannus1,HS1,HS2}. An special type of adiabatic pulse is the so-called `rapid adiabatic passage' pulse, which was
employed in trapped ions to measure the motional ground state~\cite{Gebert2016}, to control the motional states using
sideband excitation~\cite{Watanabe2011}, and to prepare Dicke states~\cite{Linington2008, Toyoda2011}.

\section{Discussion and Conclusion}
\label{sec:conclusion}

After many years of intense progress in experiments, one of the most pressing challenges of quantum computation nowadays is the enlargement of the Hilbert space available. In the CZ paradigm of ion trap quantum computing, extending the Hilbert space requires adding ever more ions to the quantum processor. As the number of ions increases, so does the influence of the environment in the form of natural decay and stray magnetic fields, the most relevant sources of environmentally driven decoherence in this kind of system. It may be said that, for general experimental realizations of a quantum processor, the desired enlargement of the Hilbert space implies as penalty greatly enlarging the \textit{physical} size of the quantum processor, and thus the volume of actual space, i.e. the `size' of the environment, probed by the quantum system. 

In this paper, we have tried to pursue a different route to enlarge the quantum configuration space, by using the volume in actual physical space sparingly: a more feasible quantum computer might be that which packs a large configuration space in a small physical volume. We have here followed an alternative route that could help to mitigate the scalability problem of the ion trap quantum processor up to a certain point. The motional degrees of freedom of a single trapped ion offer, in principle, a configuration space with dimensionality restricted only by non-linearities of the trap harmonic potential. This approach has the practical advantage of causing only modest increase (polynomial) of the physical volume occupied by the atomic wave function employed in the computation. 

The main advantage of the ion trap CVQC approach might be the ability to harness sectors of the computational configuration space which are mostly disregarded in the CZ paradigm. The CVQC approach could, in principle, allow for a substantial increase in the size of the Hilbert space available for quantum computing while using current ion trap technology. The increase in manipulation and measurement complexity implied by our scheme would not be particular to it, but rather a common trait to any actual implementation of quantum computing. The fact that our scheme seems to put those challenges within the grasp of current technology, and hence bring them to our minds as urgent matters, should be seen not only as a positive trait of our proposal, but also as a reminder of the daunting endeavor entailed in building a working quantum computer.

\section*{Acknowledgments}

This work was supported by project 473847/2012-4, funded by the Conselho Nacional de Desenvolvimento \linebreak Cient\'\i{}fico e Tecnol\'ogico (CNPq) and by the Instituto Nacional de Ci\^encia e Tecnologia de Informa\c{c}\~ao Qu\^antica (INCT-IQ). JGF thanks the Brazilian funding agency \linebreak CNPq (Grant No. BJT 300121/2015-6).

\section*{References}
\bibliography{Bibliography_CVgates6_jeff_refbib}

\begin{thebibliography}{86}
\expandafter\ifx\csname natexlab\endcsname\relax\def\natexlab#1{#1}\fi
\providecommand{\url}[1]{\texttt{#1}}
\providecommand{\href}[2]{#2}
\providecommand{\path}[1]{#1}
\providecommand{\DOIprefix}{doi:}
\providecommand{\ArXivprefix}{arXiv:}
\providecommand{\URLprefix}{URL: }
\providecommand{\Pubmedprefix}{pmid:}
\providecommand{\doi}[1]{\href{http://dx.doi.org/#1}{\path{#1}}}
\providecommand{\Pubmed}[1]{\href{pmid:#1}{\path{#1}}}
\providecommand{\bibinfo}[2]{#2}
\ifx\xfnm\relax \def\xfnm[#1]{\unskip,\space#1}\fi
\bibitem[{Nielsen and Chuang(2000)}]{nielsenchuangQCQI}
\bibinfo{author}{M.~A. Nielsen}, \bibinfo{author}{I.~L. Chuang},
  \bibinfo{title}{Quantum Computation and Quantum Information},
  \bibinfo{publisher}{Cambridge University Press}, \bibinfo{year}{2000}.
\bibitem[{Monroe et~al.(1995)Monroe, Meekhof, King, Itano, and
  Wineland}]{winelandCNOT_prl95}
\bibinfo{author}{C.~Monroe}, \bibinfo{author}{D.~M. Meekhof},
  \bibinfo{author}{B.~E. King}, \bibinfo{author}{W.~M. Itano},
  \bibinfo{author}{D.~J. Wineland}, \bibinfo{journal}{Phys. Rev. Lett.}
  \bibinfo{volume}{75} (\bibinfo{year}{1995}) \bibinfo{pages}{4714--4717}.
  \URLprefix \url{http://link.aps.org/doi/10.1103/PhysRevLett.75.4714}.
  \DOIprefix\doi{10.1103/PhysRevLett.75.4714}.
\bibitem[{Leibfried et~al.(2003)Leibfried, DeMarco, Meyer, Lucas, Barrett,
  Britton, Itano, Jelenkovic, Langer, Rosenband, and
  Wineland}]{phasegate_nature03}
\bibinfo{author}{D.~Leibfried}, \bibinfo{author}{B.~DeMarco},
  \bibinfo{author}{V.~Meyer}, \bibinfo{author}{D.~Lucas},
  \bibinfo{author}{M.~Barrett}, \bibinfo{author}{J.~Britton},
  \bibinfo{author}{W.~M. Itano}, \bibinfo{author}{B.~Jelenkovic},
  \bibinfo{author}{C.~Langer}, \bibinfo{author}{T.~Rosenband},
  \bibinfo{author}{D.~J. Wineland}, \bibinfo{journal}{Nature (London)}
  \bibinfo{volume}{422} (\bibinfo{year}{2003}) \bibinfo{pages}{412}. \URLprefix
  \url{http://dx.doi.org/10.1038/nature01492}.
  \DOIprefix\doi{doi:10.1038/nature01492}.
\bibitem[{Schmidt-Kaler et~al.(2003)Schmidt-Kaler, Haffner, Riebe, Gulde,
  Lancaster, Deuschle, Becher, Roos, Eschner, and Blatt}]{cnot_nature03}
\bibinfo{author}{F.~Schmidt-Kaler}, \bibinfo{author}{H.~Haffner},
  \bibinfo{author}{M.~Riebe}, \bibinfo{author}{S.~Gulde},
  \bibinfo{author}{G.~P.~T. Lancaster}, \bibinfo{author}{T.~Deuschle},
  \bibinfo{author}{C.~Becher}, \bibinfo{author}{C.~F. Roos},
  \bibinfo{author}{J.~Eschner}, \bibinfo{author}{R.~Blatt},
  \bibinfo{journal}{Nature (London)} \bibinfo{volume}{422}
  (\bibinfo{year}{2003}) \bibinfo{pages}{408}. \URLprefix
  \url{http://dx.doi.org/10.1038/nature01494}.
  \DOIprefix\doi{doi:10.1038/nature01494}.
\bibitem[{Cirac and Zoller(2000)}]{CZscalable_nature00}
\bibinfo{author}{J.~I. Cirac}, \bibinfo{author}{P.~Zoller},
  \bibinfo{journal}{Nature (London)} \bibinfo{volume}{404}
  (\bibinfo{year}{2000}) \bibinfo{pages}{579}. \URLprefix
  \url{http://dx.doi.org/10.1038/35007021}.
  \DOIprefix\doi{doi:10.1038/35007021}.
\bibitem[{Kielpinski et~al.(2002)Kielpinski, Monroe, and
  Wineland}]{scalabletrap_nature02}
\bibinfo{author}{D.~Kielpinski}, \bibinfo{author}{C.~Monroe},
  \bibinfo{author}{D.~J. Wineland}, \bibinfo{journal}{Nature (London)}
  \bibinfo{volume}{417} (\bibinfo{year}{2002}) \bibinfo{pages}{709}. \URLprefix
  \url{http://dx.doi.org/10.1038/nature00784}.
  \DOIprefix\doi{doi:10.1038/nature00784}.
\bibitem[{Cirac and Zoller(1995)}]{ciraczoller95}
\bibinfo{author}{J.~I. Cirac}, \bibinfo{author}{P.~Zoller},
  \bibinfo{journal}{Phys. Rev. Lett.} \bibinfo{volume}{74}
  (\bibinfo{year}{1995}) \bibinfo{pages}{4091--4094}. \URLprefix
  \url{http://link.aps.org/doi/10.1103/PhysRevLett.74.4091}.
  \DOIprefix\doi{10.1103/PhysRevLett.74.4091}.
\bibitem[{S\o{}rensen and M\o{}lmer(1999)}]{ms_prl99}
\bibinfo{author}{A.~S\o{}rensen}, \bibinfo{author}{K.~M\o{}lmer},
  \bibinfo{journal}{Phys. Rev. Lett.} \bibinfo{volume}{82}
  (\bibinfo{year}{1999}) \bibinfo{pages}{1971--1974}. \URLprefix
  \url{http://link.aps.org/doi/10.1103/PhysRevLett.82.1971}.
  \DOIprefix\doi{10.1103/PhysRevLett.82.1971}.
\bibitem[{Milburn et~al.(2000)Milburn, Schneider, and James}]{sigmaz_00}
\bibinfo{author}{G.~Milburn}, \bibinfo{author}{S.~Schneider},
  \bibinfo{author}{D.~James}, \bibinfo{journal}{Fortschritte der Physik}
  \bibinfo{volume}{48} (\bibinfo{year}{2000}) \bibinfo{pages}{801--810}.
  \URLprefix
  \url{http://dx.doi.org/10.1002/1521-3978(200009)48:9/11<801::AID-PROP801>3.0.CO;2-1}.
\bibitem[{Mizrahi et~al.(2013)Mizrahi, Senko, Neyenhuis, Johnson, Campbell,
  Conover, and Monroe}]{ultrafastentanglementmonroe_prl13}
\bibinfo{author}{J.~Mizrahi}, \bibinfo{author}{C.~Senko},
  \bibinfo{author}{B.~Neyenhuis}, \bibinfo{author}{K.~G. Johnson},
  \bibinfo{author}{W.~C. Campbell}, \bibinfo{author}{C.~W.~S. Conover},
  \bibinfo{author}{C.~Monroe}, \bibinfo{journal}{Phys. Rev. Lett.}
  \bibinfo{volume}{110} (\bibinfo{year}{2013}) \bibinfo{pages}{203001}.
  \URLprefix \url{http://link.aps.org/doi/10.1103/PhysRevLett.110.203001}.
  \DOIprefix\doi{10.1103/PhysRevLett.110.203001}.
\bibitem[{Haljan et~al.(2005)Haljan, Brickman, Deslauriers, Lee, and
  Monroe}]{spindependentMonroe_05}
\bibinfo{author}{P.~C. Haljan}, \bibinfo{author}{K.-A. Brickman},
  \bibinfo{author}{L.~Deslauriers}, \bibinfo{author}{P.~J. Lee},
  \bibinfo{author}{C.~Monroe}, \bibinfo{journal}{Phys. Rev. Lett.}
  \bibinfo{volume}{94} (\bibinfo{year}{2005}) \bibinfo{pages}{153602}.
  \URLprefix \url{http://link.aps.org/doi/10.1103/PhysRevLett.94.153602}.
  \DOIprefix\doi{10.1103/PhysRevLett.94.153602}.
\bibitem[{Lloyd and Braunstein(1999)}]{lloydbraunstein_prl99}
\bibinfo{author}{S.~Lloyd}, \bibinfo{author}{S.~L. Braunstein},
  \bibinfo{journal}{Phys. Rev. Lett.} \bibinfo{volume}{82}
  (\bibinfo{year}{1999}) \bibinfo{pages}{1784--1787}. \URLprefix
  \url{http://link.aps.org/doi/10.1103/PhysRevLett.82.1784}.
  \DOIprefix\doi{10.1103/PhysRevLett.82.1784}.
\bibitem[{Braunstein and van Loock(2005)}]{reviewbraunsteinvanloock_rmp05}
\bibinfo{author}{S.~L. Braunstein}, \bibinfo{author}{P.~van Loock},
  \bibinfo{journal}{Rev. Mod. Phys.} \bibinfo{volume}{77}
  (\bibinfo{year}{2005}) \bibinfo{pages}{513--577}. \URLprefix
  \url{http://link.aps.org/doi/10.1103/RevModPhys.77.513}.
  \DOIprefix\doi{10.1103/RevModPhys.77.513}.
\bibitem[{Arvind et~al.(2005)Arvind, Dutta, Mukunda, and Simon}]{symplectic}
\bibinfo{author}{Arvind}, \bibinfo{author}{B.~Dutta},
  \bibinfo{author}{N.~Mukunda}, \bibinfo{author}{R.~Simon},
  \bibinfo{journal}{Pramana} \bibinfo{volume}{45} (\bibinfo{year}{2005})
  \bibinfo{pages}{471--497}. \URLprefix
  \url{http://dx.doi.org/10.1007/BF02848172}.
  \DOIprefix\doi{10.1007/BF02848172}.
\bibitem[{Eisert and Plenio(2003)}]{eisert_ijqi03}
\bibinfo{author}{J.~Eisert}, \bibinfo{author}{M.~B. Plenio},
  \bibinfo{journal}{Int. J. Quantum Inform.} \bibinfo{volume}{01}
  (\bibinfo{year}{2003}) \bibinfo{pages}{479--506}. \URLprefix
  \url{http://www.worldscientific.com/doi/abs/10.1142/S0219749903000371}.
  \DOIprefix\doi{10.1142/S0219749903000371}.
\bibitem[{Bartlett et~al.(2002)Bartlett, Sanders, Braunstein, and
  Nemoto}]{bartlett_prl02}
\bibinfo{author}{S.~D. Bartlett}, \bibinfo{author}{B.~C. Sanders},
  \bibinfo{author}{S.~L. Braunstein}, \bibinfo{author}{K.~Nemoto},
  \bibinfo{journal}{Phys. Rev. Lett.} \bibinfo{volume}{88}
  (\bibinfo{year}{2002}) \bibinfo{pages}{097904}. \URLprefix
  \url{http://link.aps.org/doi/10.1103/PhysRevLett.88.097904}.
  \DOIprefix\doi{10.1103/PhysRevLett.88.097904}.
\bibitem[{Bartlett and Sanders(2002)}]{bartlett_pra02}
\bibinfo{author}{S.~D. Bartlett}, \bibinfo{author}{B.~C. Sanders},
  \bibinfo{journal}{Phys. Rev. A} \bibinfo{volume}{65} (\bibinfo{year}{2002})
  \bibinfo{pages}{042304}. \URLprefix
  \url{http://link.aps.org/doi/10.1103/PhysRevA.65.042304}.
  \DOIprefix\doi{10.1103/PhysRevA.65.042304}.
\bibitem[{Monroe et~al.(1996)Monroe, Meekhof, King, and
  Wineland}]{wineland_science96}
\bibinfo{author}{C.~Monroe}, \bibinfo{author}{D.~M. Meekhof},
  \bibinfo{author}{B.~E. King}, \bibinfo{author}{D.~J. Wineland},
  \bibinfo{journal}{Science} \bibinfo{volume}{272} (\bibinfo{year}{1996})
  \bibinfo{pages}{1131--1136}. \URLprefix
  \url{http://science.sciencemag.org/content/272/5265/1131}.
  \DOIprefix\doi{10.1126/science.272.5265.1131}.
\bibitem[{Solano et~al.(1999)Solano, de~Matos~Filho, and Zagury}]{zagury_pra99}
\bibinfo{author}{E.~Solano}, \bibinfo{author}{R.~L. de~Matos~Filho},
  \bibinfo{author}{N.~Zagury}, \bibinfo{journal}{Phys. Rev. A}
  \bibinfo{volume}{59} (\bibinfo{year}{1999}) \bibinfo{pages}{R2539--R2543}.
  \URLprefix \url{http://link.aps.org/doi/10.1103/PhysRevA.59.R2539}.
  \DOIprefix\doi{10.1103/PhysRevA.59.R2539}.
\bibitem[{Leibfried et~al.(1996)Leibfried, Meekhof, King, Monroe, Itano, and
  Wineland}]{wineland2_prl96}
\bibinfo{author}{D.~Leibfried}, \bibinfo{author}{D.~M. Meekhof},
  \bibinfo{author}{B.~E. King}, \bibinfo{author}{C.~Monroe},
  \bibinfo{author}{W.~M. Itano}, \bibinfo{author}{D.~J. Wineland},
  \bibinfo{journal}{Phys. Rev. Lett.} \bibinfo{volume}{77}
  (\bibinfo{year}{1996}) \bibinfo{pages}{4281--4285}. \URLprefix
  \url{http://link.aps.org/doi/10.1103/PhysRevLett.77.4281}.
  \DOIprefix\doi{10.1103/PhysRevLett.77.4281}.
\bibitem[{Solano et~al.(2001)Solano, de~Matos~Filho, and
  Zagury}]{mesoscopicmotion_prl01}
\bibinfo{author}{E.~Solano}, \bibinfo{author}{R.~L. de~Matos~Filho},
  \bibinfo{author}{N.~Zagury}, \bibinfo{journal}{Phys. Rev. Lett.}
  \bibinfo{volume}{87} (\bibinfo{year}{2001}) \bibinfo{pages}{060402}.
  \URLprefix \url{http://link.aps.org/doi/10.1103/PhysRevLett.87.060402}.
  \DOIprefix\doi{10.1103/PhysRevLett.87.060402}.
\bibitem[{Vlastakis et~al.(2013)Vlastakis, Kirchmair, Leghtas, Nigg, Frunzio,
  Girvin, Mirrahimi, Devoret, and
  Schoelkopf}]{qbitenconding100photon_science13}
\bibinfo{author}{B.~Vlastakis}, \bibinfo{author}{G.~Kirchmair},
  \bibinfo{author}{Z.~Leghtas}, \bibinfo{author}{S.~E. Nigg},
  \bibinfo{author}{L.~Frunzio}, \bibinfo{author}{S.~M. Girvin},
  \bibinfo{author}{M.~Mirrahimi}, \bibinfo{author}{M.~H. Devoret},
  \bibinfo{author}{R.~J. Schoelkopf}, \bibinfo{journal}{Science}
  \bibinfo{volume}{342} (\bibinfo{year}{2013}) \bibinfo{pages}{607--610}.
  \URLprefix \url{http://science.sciencemag.org/content/342/6158/607}.
  \DOIprefix\doi{10.1126/science.1243289}.
\bibitem[{Haroche and Raimond(2006)}]{haroche_exploringthequantum}
\bibinfo{author}{S.~Haroche}, \bibinfo{author}{J.-M. Raimond},
  \bibinfo{title}{Exploring the Quantum: Atoms, Cavities, and Photons},
  \bibinfo{publisher}{Oxford Univ. Press}, \bibinfo{year}{2006}.
\bibitem[{Serafini et~al.(2009)Serafini, Retzker, and Plenio}]{Alessio2009}
\bibinfo{author}{A.~Serafini}, \bibinfo{author}{A.~Retzker},
  \bibinfo{author}{M.~B. Plenio}, \bibinfo{journal}{New J. Phys.}
  \bibinfo{volume}{11} (\bibinfo{year}{2009}) \bibinfo{pages}{023007}.
  \URLprefix \url{http://stacks.iop.org/1367-2630/11/i=2/a=023007}.
\bibitem[{Alonso et~al.(2013)Alonso, Leupold, Keitch, and Home}]{Alonso2013}
\bibinfo{author}{J.~Alonso}, \bibinfo{author}{F.~M. Leupold},
  \bibinfo{author}{B.~C. Keitch}, \bibinfo{author}{J.~P. Home},
  \bibinfo{journal}{New J. Phys.} \bibinfo{volume}{15} (\bibinfo{year}{2013})
  \bibinfo{pages}{023001}. \URLprefix
  \url{http://stacks.iop.org/1367-2630/15/i=2/a=023001}.
\bibitem[{Ding et~al.(2014)Ding, Loh, Hablutzel, Gao, Maslennikov, and
  Matsukevich}]{Ding2014}
\bibinfo{author}{S.~Ding}, \bibinfo{author}{H.~Loh},
  \bibinfo{author}{R.~Hablutzel}, \bibinfo{author}{M.~Gao},
  \bibinfo{author}{G.~Maslennikov}, \bibinfo{author}{D.~Matsukevich},
  \bibinfo{journal}{Phys. Rev. Lett.} \bibinfo{volume}{113}
  (\bibinfo{year}{2014}) \bibinfo{pages}{073002}. \URLprefix
  \url{http://link.aps.org/doi/10.1103/PhysRevLett.113.073002}.
  \DOIprefix\doi{10.1103/PhysRevLett.113.073002}.
\bibitem[{Mirkhalaf and M\o{}lmer(2012)}]{Mirkhalaf2012}
\bibinfo{author}{S.~Mirkhalaf}, \bibinfo{author}{K.~M\o{}lmer},
  \bibinfo{journal}{Phys. Rev. A} \bibinfo{volume}{85} (\bibinfo{year}{2012})
  \bibinfo{pages}{042109}. \URLprefix
  \url{http://link.aps.org/doi/10.1103/PhysRevA.85.042109}.
  \DOIprefix\doi{10.1103/PhysRevA.85.042109}.
\bibitem[{Johnson et~al.(2015)Johnson, Neyenhuis, Mizrahi, Wong-Campos, and
  Monroe}]{Johnson2015}
\bibinfo{author}{K.~G. Johnson}, \bibinfo{author}{B.~Neyenhuis},
  \bibinfo{author}{J.~Mizrahi}, \bibinfo{author}{J.~D. Wong-Campos},
  \bibinfo{author}{C.~Monroe}, \bibinfo{journal}{Phys. Rev. Lett.}
  \bibinfo{volume}{115} (\bibinfo{year}{2015}) \bibinfo{pages}{213001}.
  \URLprefix \url{http://link.aps.org/doi/10.1103/PhysRevLett.115.213001}.
  \DOIprefix\doi{10.1103/PhysRevLett.115.213001}.
\bibitem[{Hashemloo et~al.(2016)Hashemloo, Dion, and Rahali}]{Hashemloo2016}
\bibinfo{author}{A.~Hashemloo}, \bibinfo{author}{C.~M. Dion},
  \bibinfo{author}{G.~Rahali}, \bibinfo{journal}{Int. J. Mod. Phys. C}
  \bibinfo{volume}{27} (\bibinfo{year}{2016}) \bibinfo{pages}{1650014}.
  \URLprefix
  \url{http://www.worldscientific.com/doi/abs/10.1142/S0129183116500145}.
  \DOIprefix\doi{10.1142/S0129183116500145}.
\bibitem[{Nicacio et~al.(2013)Nicacio, Furuya, and Semi\~ao}]{Nicacio2013}
\bibinfo{author}{F.~Nicacio}, \bibinfo{author}{K.~Furuya},
  \bibinfo{author}{F.~L. Semi\~ao}, \bibinfo{journal}{Phys. Rev. A}
  \bibinfo{volume}{88} (\bibinfo{year}{2013}) \bibinfo{pages}{022330}.
  \URLprefix \url{http://link.aps.org/doi/10.1103/PhysRevA.88.022330}.
  \DOIprefix\doi{10.1103/PhysRevA.88.022330}.
\bibitem[{Xu et~al.(2013)Xu, Wang, Zhang, and Liu}]{Xu2013b}
\bibinfo{author}{X.-W. Xu}, \bibinfo{author}{H.~Wang},
  \bibinfo{author}{J.~Zhang}, \bibinfo{author}{Y.-x. Liu},
  \bibinfo{journal}{Phys. Rev. A} \bibinfo{volume}{88} (\bibinfo{year}{2013})
  \bibinfo{pages}{063819}. \URLprefix
  \url{http://link.aps.org/doi/10.1103/PhysRevA.88.063819}.
  \DOIprefix\doi{10.1103/PhysRevA.88.063819}.
\bibitem[{Miry and Tavassoly(2012)}]{Miry2012}
\bibinfo{author}{S.~R. Miry}, \bibinfo{author}{M.~K. Tavassoly},
  \bibinfo{journal}{J. Phys. B} \bibinfo{volume}{45} (\bibinfo{year}{2012})
  \bibinfo{pages}{175502}. \URLprefix
  \url{http://stacks.iop.org/0953-4075/45/i=17/a=175502}.
\bibitem[{Rodr{\'i}guez-M{\'e}ndez and Moya-Cessa(2012)}]{Roriguez2012}
\bibinfo{author}{D.~Rodr{\'i}guez-M{\'e}ndez}, \bibinfo{author}{H.~Moya-Cessa},
  \bibinfo{journal}{Phys. Scr.} \bibinfo{volume}{2012} (\bibinfo{year}{2012})
  \bibinfo{pages}{014028}. \URLprefix
  \url{http://stacks.iop.org/1402-4896/2012/i=T147/a=014028}.
\bibitem[{Haze et~al.(2012)Haze, Tateishi, Noguchi, Toyoda, and
  Urabe}]{Haze2012}
\bibinfo{author}{S.~Haze}, \bibinfo{author}{Y.~Tateishi},
  \bibinfo{author}{A.~Noguchi}, \bibinfo{author}{K.~Toyoda},
  \bibinfo{author}{S.~Urabe}, \bibinfo{journal}{Phys. Rev. A}
  \bibinfo{volume}{85} (\bibinfo{year}{2012}) \bibinfo{pages}{031401}.
  \URLprefix \url{http://link.aps.org/doi/10.1103/PhysRevA.85.031401}.
  \DOIprefix\doi{10.1103/PhysRevA.85.031401}.
\bibitem[{Dutta et~al.(2012)Dutta, Mukherjee, and Sengupta}]{Dutta2012}
\bibinfo{author}{T.~Dutta}, \bibinfo{author}{M.~Mukherjee},
  \bibinfo{author}{K.~Sengupta}, \bibinfo{journal}{Phys. Rev. A}
  \bibinfo{volume}{85} (\bibinfo{year}{2012}) \bibinfo{pages}{063401}.
  \URLprefix \url{http://link.aps.org/doi/10.1103/PhysRevA.85.063401}.
  \DOIprefix\doi{10.1103/PhysRevA.85.063401}.
\bibitem[{Dutta et~al.(2013)Dutta, Mukherjee, and Sengupta}]{Dutta2013}
\bibinfo{author}{T.~Dutta}, \bibinfo{author}{M.~Mukherjee},
  \bibinfo{author}{K.~Sengupta}, \bibinfo{journal}{Phys. Rev. Lett.}
  \bibinfo{volume}{111} (\bibinfo{year}{2013}) \bibinfo{pages}{170406}.
  \URLprefix \url{http://link.aps.org/doi/10.1103/PhysRevLett.111.170406}.
  \DOIprefix\doi{10.1103/PhysRevLett.111.170406}.
\bibitem[{Orszag and Larrain(2005)}]{Orszag2005}
\bibinfo{author}{M.~Orszag}, \bibinfo{author}{F.~Larrain}, \bibinfo{journal}{J.
  Opt. B: Quant. Semiclass. Opt.} \bibinfo{volume}{7} (\bibinfo{year}{2005})
  \bibinfo{pages}{S754}. \URLprefix
  \url{http://stacks.iop.org/1464-4266/7/i=12/a=045}.
\bibitem[{Shi-Jun et~al.(2008)Shi-Jun, Chi, Wen-Hai, and Liu}]{Zhang2008}
\bibinfo{author}{Z.~Shi-Jun}, \bibinfo{author}{M.~Chi},
  \bibinfo{author}{Z.~Wen-Hai}, \bibinfo{author}{Y.~Liu},
  \bibinfo{journal}{Chin. Phys. B} \bibinfo{volume}{17} (\bibinfo{year}{2008})
  \bibinfo{pages}{1593}. \DOIprefix\doi{10.1088/1674-1056/17/5/010}.
\bibitem[{Lau and James(2012)}]{Lau2012}
\bibinfo{author}{H.-K. Lau}, \bibinfo{author}{D.~F.~V. James},
  \bibinfo{journal}{Phys. Rev. A} \bibinfo{volume}{85} (\bibinfo{year}{2012})
  \bibinfo{pages}{062329}. \URLprefix
  \url{http://link.aps.org/doi/10.1103/PhysRevA.85.062329}.
  \DOIprefix\doi{10.1103/PhysRevA.85.062329}.
\bibitem[{de~Matos~Filho and Vogel(1996)}]{matosfilhovogel_prl96}
\bibinfo{author}{R.~L. de~Matos~Filho}, \bibinfo{author}{W.~Vogel},
  \bibinfo{journal}{Phys. Rev. Lett.} \bibinfo{volume}{76}
  (\bibinfo{year}{1996}) \bibinfo{pages}{4520--4523}. \URLprefix
  \url{http://link.aps.org/doi/10.1103/PhysRevLett.76.4520}.
  \DOIprefix\doi{10.1103/PhysRevLett.76.4520}.
\bibitem[{Davidovich et~al.(1996)Davidovich, Orszag, and
  Zagury}]{davidovich_pra96}
\bibinfo{author}{L.~Davidovich}, \bibinfo{author}{M.~Orszag},
  \bibinfo{author}{N.~Zagury}, \bibinfo{journal}{Phys. Rev. A}
  \bibinfo{volume}{54} (\bibinfo{year}{1996}) \bibinfo{pages}{5118--5125}.
  \URLprefix \url{http://link.aps.org/doi/10.1103/PhysRevA.54.5118}.
  \DOIprefix\doi{10.1103/PhysRevA.54.5118}.
\bibitem[{Lutterbach and Davidovich(1997)}]{lutterbachdavidovich_prl97}
\bibinfo{author}{L.~G. Lutterbach}, \bibinfo{author}{L.~Davidovich},
  \bibinfo{journal}{Phys. Rev. Lett.} \bibinfo{volume}{78}
  (\bibinfo{year}{1997}) \bibinfo{pages}{2547--2550}. \URLprefix
  \url{http://link.aps.org/doi/10.1103/PhysRevLett.78.2547}.
  \DOIprefix\doi{10.1103/PhysRevLett.78.2547}.
\bibitem[{S\o{}rensen and M\o{}lmer(2000)}]{ms_pra00}
\bibinfo{author}{A.~S\o{}rensen}, \bibinfo{author}{K.~M\o{}lmer},
  \bibinfo{journal}{Phys. Rev. A} \bibinfo{volume}{62} (\bibinfo{year}{2000})
  \bibinfo{pages}{022311}. \URLprefix
  \url{http://link.aps.org/doi/10.1103/PhysRevA.62.022311}.
  \DOIprefix\doi{10.1103/PhysRevA.62.022311}.
\bibitem[{Roos(2008)}]{roos_njp08}
\bibinfo{author}{C.~F. Roos}, \bibinfo{journal}{New J. Phys.}
  \bibinfo{volume}{10} (\bibinfo{year}{2008}) \bibinfo{pages}{013002}.
  \URLprefix \url{http://stacks.iop.org/1367-2630/10/i=1/a=013002}.
\bibitem[{Meekhof et~al.(1996)Meekhof, Monroe, King, Itano, and
  Wineland}]{wineland_prl96}
\bibinfo{author}{D.~M. Meekhof}, \bibinfo{author}{C.~Monroe},
  \bibinfo{author}{B.~E. King}, \bibinfo{author}{W.~M. Itano},
  \bibinfo{author}{D.~J. Wineland}, \bibinfo{journal}{Phys. Rev. Lett.}
  \bibinfo{volume}{76} (\bibinfo{year}{1996}) \bibinfo{pages}{1796--1799}.
  \URLprefix \url{http://link.aps.org/doi/10.1103/PhysRevLett.76.1796}.
  \DOIprefix\doi{10.1103/PhysRevLett.76.1796}.
\bibitem[{Lo et~al.(2015)Lo, Kienzler, de~Clercq, Marinelli, Negnevitsky,
  Keitch, and Home}]{squeezingHome_nature15}
\bibinfo{author}{H.-Y. Lo}, \bibinfo{author}{D.~Kienzler},
  \bibinfo{author}{L.~de~Clercq}, \bibinfo{author}{M.~Marinelli},
  \bibinfo{author}{V.~Negnevitsky}, \bibinfo{author}{B.~C. Keitch},
  \bibinfo{author}{J.~P. Home}, \bibinfo{journal}{Nature (London)}
  \bibinfo{volume}{521} (\bibinfo{year}{2015}) \bibinfo{pages}{336}. \URLprefix
  \url{http://dx.doi.org/10.1038/nature14458}.
  \DOIprefix\doi{doi:10.1038/nature14458}.
\bibitem[{Leibfried et~al.(2003)Leibfried, Blatt, Monroe, and
  Wineland}]{rmpiontrap_03}
\bibinfo{author}{D.~Leibfried}, \bibinfo{author}{R.~Blatt},
  \bibinfo{author}{C.~Monroe}, \bibinfo{author}{D.~Wineland},
  \bibinfo{journal}{Rev. Mod. Phys.} \bibinfo{volume}{75}
  (\bibinfo{year}{2003}) \bibinfo{pages}{281--324}. \URLprefix
  \url{http://link.aps.org/doi/10.1103/RevModPhys.75.281}.
  \DOIprefix\doi{10.1103/RevModPhys.75.281}.
\bibitem[{Roghani and Helm(2008)}]{Roghani}
\bibinfo{author}{M.~Roghani}, \bibinfo{author}{H.~Helm},
  \bibinfo{journal}{Phys. Rev. A} \bibinfo{volume}{77} (\bibinfo{year}{2008})
  \bibinfo{pages}{043418}. \URLprefix
  \url{https://link.aps.org/doi/10.1103/PhysRevA.77.043418}.
  \DOIprefix\doi{10.1103/PhysRevA.77.043418}.
\bibitem[{Moya-Cessa et~al.(2012)Moya-Cessa, Soto-Eguibar, Vargas-Martínez,
  Juárez-Amaro, and Zúñiga-Segundo}]{Moya}
\bibinfo{author}{H.~Moya-Cessa}, \bibinfo{author}{F.~Soto-Eguibar},
  \bibinfo{author}{J.~M. Vargas-Martínez}, \bibinfo{author}{R.~Juárez-Amaro},
  \bibinfo{author}{A.~Zúñiga-Segundo}, \bibinfo{journal}{Physics Reports}
  \bibinfo{volume}{513} (\bibinfo{year}{2012}) \bibinfo{pages}{229 -- 261}.
  \URLprefix
  \url{http://www.sciencedirect.com/science/article/pii/S0370157312000117}.
  \DOIprefix\doi{http://dx.doi.org/10.1016/j.physrep.2012.01.002},
  \bibinfo{note}{ion-laser interactions: the most complete solution}.
\bibitem[{Einstein et~al.(1935)Einstein, Podolsky, and Rosen}]{epr_pr35}
\bibinfo{author}{A.~Einstein}, \bibinfo{author}{B.~Podolsky},
  \bibinfo{author}{N.~Rosen}, \bibinfo{journal}{Phys. Rev.}
  \bibinfo{volume}{47} (\bibinfo{year}{1935}) \bibinfo{pages}{777--780}.
  \URLprefix \url{http://link.aps.org/doi/10.1103/PhysRev.47.777}.
  \DOIprefix\doi{10.1103/PhysRev.47.777}.
\bibitem[{Jefferts et~al.(1995)Jefferts, Monroe, Bell, and
  Wineland}]{winelandtrapasym_pra95}
\bibinfo{author}{S.~R. Jefferts}, \bibinfo{author}{C.~Monroe},
  \bibinfo{author}{E.~W. Bell}, \bibinfo{author}{D.~J. Wineland},
  \bibinfo{journal}{Phys. Rev. A} \bibinfo{volume}{51} (\bibinfo{year}{1995})
  \bibinfo{pages}{3112--3116}. \URLprefix
  \url{http://link.aps.org/doi/10.1103/PhysRevA.51.3112}.
  \DOIprefix\doi{10.1103/PhysRevA.51.3112}.
\bibitem[{Wineland et~al.(1998)Wineland, Monroe, Itano, Leibfried, King, and
  Meekhof}]{Wineland:1997mg}
\bibinfo{author}{D.~J. Wineland}, \bibinfo{author}{C.~Monroe},
  \bibinfo{author}{W.~M. Itano}, \bibinfo{author}{D.~Leibfried},
  \bibinfo{author}{B.~E. King}, \bibinfo{author}{D.~M. Meekhof},
  \bibinfo{journal}{J. Res. Natl. Inst. Stand. Tech.} \bibinfo{volume}{103}
  (\bibinfo{year}{1998}) \bibinfo{pages}{259}.
  \href{http://arxiv.org/abs/quant-ph/9710025}{\tt arXiv:quant-ph/9710025}.
\bibitem[{Poschinger et~al.(2012)Poschinger, Walther, Hettrich, Ziesel, and
  Schmidt-Kaler}]{Poschinger2012}
\bibinfo{author}{U.~Poschinger}, \bibinfo{author}{A.~Walther},
  \bibinfo{author}{M.~Hettrich}, \bibinfo{author}{F.~Ziesel},
  \bibinfo{author}{F.~Schmidt-Kaler}, \bibinfo{journal}{Applied Physics B}
  \bibinfo{volume}{107} (\bibinfo{year}{2012}) \bibinfo{pages}{1159--1165}.
  \URLprefix \url{http://dx.doi.org/10.1007/s00340-012-4882-3}.
  \DOIprefix\doi{10.1007/s00340-012-4882-3}.
\bibitem[{Z\"ahringer et~al.(2010)Z\"ahringer, Kirchmair, Gerritsma, Solano,
  Blatt, and Roos}]{quantumwalkions_10}
\bibinfo{author}{F.~Z\"ahringer}, \bibinfo{author}{G.~Kirchmair},
  \bibinfo{author}{R.~Gerritsma}, \bibinfo{author}{E.~Solano},
  \bibinfo{author}{R.~Blatt}, \bibinfo{author}{C.~F. Roos},
  \bibinfo{journal}{Phys. Rev. Lett.} \bibinfo{volume}{104}
  (\bibinfo{year}{2010}) \bibinfo{pages}{100503}. \URLprefix
  \url{http://link.aps.org/doi/10.1103/PhysRevLett.104.100503}.
  \DOIprefix\doi{10.1103/PhysRevLett.104.100503}.
\bibitem[{Monz et~al.(2011)Monz, Schindler, Barreiro, Chwalla, Nigg, Coish,
  Harlander, H\"ansel, Hennrich, and Blatt}]{qubits14_prl11}
\bibinfo{author}{T.~Monz}, \bibinfo{author}{P.~Schindler},
  \bibinfo{author}{J.~T. Barreiro}, \bibinfo{author}{M.~Chwalla},
  \bibinfo{author}{D.~Nigg}, \bibinfo{author}{W.~A. Coish},
  \bibinfo{author}{M.~Harlander}, \bibinfo{author}{W.~H\"ansel},
  \bibinfo{author}{M.~Hennrich}, \bibinfo{author}{R.~Blatt},
  \bibinfo{journal}{Phys. Rev. Lett.} \bibinfo{volume}{106}
  (\bibinfo{year}{2011}) \bibinfo{pages}{130506}. \URLprefix
  \url{http://link.aps.org/doi/10.1103/PhysRevLett.106.130506}.
  \DOIprefix\doi{10.1103/PhysRevLett.106.130506}.
\bibitem[{Lloyd and Slotine(1998)}]{cverrorcorr_prl98}
\bibinfo{author}{S.~Lloyd}, \bibinfo{author}{J.-J.~E. Slotine},
  \bibinfo{journal}{Phys. Rev. Lett.} \bibinfo{volume}{80}
  (\bibinfo{year}{1998}) \bibinfo{pages}{4088--4091}. \URLprefix
  \url{http://link.aps.org/doi/10.1103/PhysRevLett.80.4088}.
  \DOIprefix\doi{10.1103/PhysRevLett.80.4088}.
\bibitem[{Abrams and Lloyd(1999)}]{lloyd_prl99}
\bibinfo{author}{D.~S. Abrams}, \bibinfo{author}{S.~Lloyd},
  \bibinfo{journal}{Phys. Rev. Lett.} \bibinfo{volume}{83}
  (\bibinfo{year}{1999}) \bibinfo{pages}{5162--5165}. \URLprefix
  \url{http://link.aps.org/doi/10.1103/PhysRevLett.83.5162}.
  \DOIprefix\doi{10.1103/PhysRevLett.83.5162}.
\bibitem[{Gottesman et~al.(2001)Gottesman, Kitaev, and
  Preskill}]{gottesmanencoding_pra01}
\bibinfo{author}{D.~Gottesman}, \bibinfo{author}{A.~Kitaev},
  \bibinfo{author}{J.~Preskill}, \bibinfo{journal}{Phys. Rev. A}
  \bibinfo{volume}{64} (\bibinfo{year}{2001}) \bibinfo{pages}{012310}.
  \URLprefix \url{http://link.aps.org/doi/10.1103/PhysRevA.64.012310}.
  \DOIprefix\doi{10.1103/PhysRevA.64.012310}.
\bibitem[{Schmidt-Kaler et~al.(2003)Schmidt-Kaler, Gulde, Riebe, Deuschle,
  Kreuter, Lancaster, Becher, Eschner, Häffner, and Blatt}]{Schmidt-Kaler2003}
\bibinfo{author}{F.~Schmidt-Kaler}, \bibinfo{author}{S.~Gulde},
  \bibinfo{author}{M.~Riebe}, \bibinfo{author}{T.~Deuschle},
  \bibinfo{author}{A.~Kreuter}, \bibinfo{author}{G.~Lancaster},
  \bibinfo{author}{C.~Becher}, \bibinfo{author}{J.~Eschner},
  \bibinfo{author}{H.~Häffner}, \bibinfo{author}{R.~Blatt},
  \bibinfo{journal}{Journal of Physics B: Atomic, Molecular and Optical
  Physics} \bibinfo{volume}{36} (\bibinfo{year}{2003}) \bibinfo{pages}{623}.
  \URLprefix \url{http://stacks.iop.org/0953-4075/36/i=3/a=319}.
\bibitem[{Monz et~al.(2009)Monz, Kim, H\"ansel, Riebe, Villar, Schindler,
  Chwalla, Hennrich, and Blatt}]{Monz2009}
\bibinfo{author}{T.~Monz}, \bibinfo{author}{K.~Kim},
  \bibinfo{author}{W.~H\"ansel}, \bibinfo{author}{M.~Riebe},
  \bibinfo{author}{A.~S. Villar}, \bibinfo{author}{P.~Schindler},
  \bibinfo{author}{M.~Chwalla}, \bibinfo{author}{M.~Hennrich},
  \bibinfo{author}{R.~Blatt}, \bibinfo{journal}{Phys. Rev. Lett.}
  \bibinfo{volume}{102} (\bibinfo{year}{2009}) \bibinfo{pages}{040501}.
  \URLprefix \url{http://link.aps.org/doi/10.1103/PhysRevLett.102.040501}.
  \DOIprefix\doi{10.1103/PhysRevLett.102.040501}.
\bibitem[{Lucas et~al.(2007)Lucas, Keitch, Home, Imreh, McDonnell, Stacey,
  Szwer, and Steane}]{Lucas2007}
\bibinfo{author}{D.~M. Lucas}, \bibinfo{author}{B.~C. Keitch},
  \bibinfo{author}{J.~P. Home}, \bibinfo{author}{G.~Imreh},
  \bibinfo{author}{M.~J. McDonnell}, \bibinfo{author}{D.~N. Stacey},
  \bibinfo{author}{D.~J. Szwer}, \bibinfo{author}{A.~M. Steane},
  \bibinfo{title}{A long-lived memory qubit on a low-decoherence quantum bus},
  \bibinfo{year}{2007}. \href{http://arxiv.org/abs/arXiv:0710.4421}{\tt
  arXiv:arXiv:0710.4421}.
\bibitem[{Roos(2008)}]{Roos2008}
\bibinfo{author}{C.~F. Roos}, \bibinfo{journal}{New Journal of Physics}
  \bibinfo{volume}{10} (\bibinfo{year}{2008}) \bibinfo{pages}{013002}.
  \URLprefix \url{http://stacks.iop.org/1367-2630/10/i=1/a=013002}.
\bibitem[{Li and Benjamin(2016)}]{Li2016}
\bibinfo{author}{Y.~Li}, \bibinfo{author}{S.~C. Benjamin},
  \bibinfo{journal}{Phys. Rev. A} \bibinfo{volume}{94} (\bibinfo{year}{2016})
  \bibinfo{pages}{042303}. \URLprefix
  \url{http://link.aps.org/doi/10.1103/PhysRevA.94.042303}.
  \DOIprefix\doi{10.1103/PhysRevA.94.042303}.
\bibitem[{Chuang et~al.(1998)Chuang, Vandersypen, Zhou, Leung, and
  Lloyd}]{nmrQuantumAlgorithm_98}
\bibinfo{author}{I.~L. Chuang}, \bibinfo{author}{L.~M.~K. Vandersypen},
  \bibinfo{author}{X.~Zhou}, \bibinfo{author}{D.~W. Leung},
  \bibinfo{author}{S.~Lloyd}, \bibinfo{journal}{Nature (London)}
  \bibinfo{volume}{393} (\bibinfo{year}{1998}) \bibinfo{pages}{143}. \URLprefix
  \url{http://dx.doi.org/10.1038/30181}. \DOIprefix\doi{doi:10.1038/30181}.
\bibitem[{Gulde et~al.(2003)Gulde, Riebe, Lancaster, Becher, Eschner, Haffner,
  Schmidt-Kaler, Chuang, and Blatt}]{jozsaBlatt_nature02}
\bibinfo{author}{S.~Gulde}, \bibinfo{author}{M.~Riebe},
  \bibinfo{author}{G.~P.~T. Lancaster}, \bibinfo{author}{C.~Becher},
  \bibinfo{author}{J.~Eschner}, \bibinfo{author}{H.~Haffner},
  \bibinfo{author}{F.~Schmidt-Kaler}, \bibinfo{author}{I.~L. Chuang},
  \bibinfo{author}{R.~Blatt}, \bibinfo{journal}{Nature (London)}
  \bibinfo{volume}{421} (\bibinfo{year}{2003}) \bibinfo{pages}{48}. \URLprefix
  \url{http://dx.doi.org/10.1038/nature01336}.
  \DOIprefix\doi{doi:10.1038/nature01336}.
\bibitem[{Chiaverini et~al.(2005)Chiaverini, Britton, Leibfried, Knill,
  Barrett, Blakestad, Itano, Jost, Langer, Ozeri, Schaetz, and
  Wineland}]{fourierWineland_science05}
\bibinfo{author}{J.~Chiaverini}, \bibinfo{author}{J.~Britton},
  \bibinfo{author}{D.~Leibfried}, \bibinfo{author}{E.~Knill},
  \bibinfo{author}{M.~D. Barrett}, \bibinfo{author}{R.~B. Blakestad},
  \bibinfo{author}{W.~M. Itano}, \bibinfo{author}{J.~D. Jost},
  \bibinfo{author}{C.~Langer}, \bibinfo{author}{R.~Ozeri},
  \bibinfo{author}{T.~Schaetz}, \bibinfo{author}{D.~J. Wineland},
  \bibinfo{journal}{Science} \bibinfo{volume}{308} (\bibinfo{year}{2005})
  \bibinfo{pages}{997--1000}. \URLprefix
  \url{http://science.sciencemag.org/content/308/5724/997}.
  \DOIprefix\doi{10.1126/science.1110335}.
\bibitem[{Genoni et~al.(2013)Genoni, Palma, Tufarelli, Olivares, Kim, and
  Paris}]{paris_pra31}
\bibinfo{author}{M.~G. Genoni}, \bibinfo{author}{M.~L. Palma},
  \bibinfo{author}{T.~Tufarelli}, \bibinfo{author}{S.~Olivares},
  \bibinfo{author}{M.~S. Kim}, \bibinfo{author}{M.~G.~A. Paris},
  \bibinfo{journal}{Phys. Rev. A} \bibinfo{volume}{87} (\bibinfo{year}{2013})
  \bibinfo{pages}{062104}. \URLprefix
  \url{http://link.aps.org/doi/10.1103/PhysRevA.87.062104}.
  \DOIprefix\doi{10.1103/PhysRevA.87.062104}.
\bibitem[{Rahimi-Keshari et~al.(2016)Rahimi-Keshari, Ralph, and
  Caves}]{cavesralphbosonsampling2016}
\bibinfo{author}{S.~Rahimi-Keshari}, \bibinfo{author}{T.~C. Ralph},
  \bibinfo{author}{C.~M. Caves}, \bibinfo{journal}{Phys. Rev. X}
  \bibinfo{volume}{6} (\bibinfo{year}{2016}) \bibinfo{pages}{021039}.
  \URLprefix \url{http://link.aps.org/doi/10.1103/PhysRevX.6.021039}.
  \DOIprefix\doi{10.1103/PhysRevX.6.021039}.
\bibitem[{Schmidt-Kaler et~al.(2003)Schmidt-Kaler, H{\"a}ffner, Gulde, Riebe,
  Lancaster, Deuschle, Becher, H{\"a}nsel, Eschner, Roos, and
  Blatt}]{trapinnsbruck_apb03}
\bibinfo{author}{F.~Schmidt-Kaler}, \bibinfo{author}{H.~H{\"a}ffner},
  \bibinfo{author}{S.~Gulde}, \bibinfo{author}{M.~Riebe},
  \bibinfo{author}{G.~Lancaster}, \bibinfo{author}{T.~Deuschle},
  \bibinfo{author}{C.~Becher}, \bibinfo{author}{W.~H{\"a}nsel},
  \bibinfo{author}{J.~Eschner}, \bibinfo{author}{C.~Roos},
  \bibinfo{author}{R.~Blatt}, \bibinfo{journal}{Appl. Phys. B}
  \bibinfo{volume}{77} (\bibinfo{year}{2003}) \bibinfo{pages}{789--796}.
  \URLprefix \url{http://dx.doi.org/10.1007/s00340-003-1346-9}.
  \DOIprefix\doi{10.1007/s00340-003-1346-9}.
\bibitem[{Campbell et~al.(2010)Campbell, Mizrahi, Quraishi, Senko, Hayes,
  Hucul, Matsukevich, Maunz, and Monroe}]{monroeultrafast_prl10}
\bibinfo{author}{W.~C. Campbell}, \bibinfo{author}{J.~Mizrahi},
  \bibinfo{author}{Q.~Quraishi}, \bibinfo{author}{C.~Senko},
  \bibinfo{author}{D.~Hayes}, \bibinfo{author}{D.~Hucul},
  \bibinfo{author}{D.~N. Matsukevich}, \bibinfo{author}{P.~Maunz},
  \bibinfo{author}{C.~Monroe}, \bibinfo{journal}{Phys. Rev. Lett.}
  \bibinfo{volume}{105} (\bibinfo{year}{2010}) \bibinfo{pages}{090502}.
  \URLprefix \url{http://link.aps.org/doi/10.1103/PhysRevLett.105.090502}.
  \DOIprefix\doi{10.1103/PhysRevLett.105.090502}.
\bibitem[{Deslauriers et~al.(2006)Deslauriers, Olmschenk, Stick, Hensinger,
  Sterk, and Monroe}]{suppressionheatingMonroe_prl06}
\bibinfo{author}{L.~Deslauriers}, \bibinfo{author}{S.~Olmschenk},
  \bibinfo{author}{D.~Stick}, \bibinfo{author}{W.~K. Hensinger},
  \bibinfo{author}{J.~Sterk}, \bibinfo{author}{C.~Monroe},
  \bibinfo{journal}{Phys. Rev. Lett.} \bibinfo{volume}{97}
  (\bibinfo{year}{2006}) \bibinfo{pages}{103007}. \URLprefix
  \url{http://link.aps.org/doi/10.1103/PhysRevLett.97.103007}.
  \DOIprefix\doi{10.1103/PhysRevLett.97.103007}.
\bibitem[{Labaziewicz et~al.(2008)Labaziewicz, Ge, Antohi, Leibrandt, Brown,
  and Chuang}]{suppressionheatingChuang_prl08}
\bibinfo{author}{J.~Labaziewicz}, \bibinfo{author}{Y.~Ge},
  \bibinfo{author}{P.~Antohi}, \bibinfo{author}{D.~Leibrandt},
  \bibinfo{author}{K.~R. Brown}, \bibinfo{author}{I.~L. Chuang},
  \bibinfo{journal}{Phys. Rev. Lett.} \bibinfo{volume}{100}
  (\bibinfo{year}{2008}) \bibinfo{pages}{013001}. \URLprefix
  \url{http://link.aps.org/doi/10.1103/PhysRevLett.100.013001}.
  \DOIprefix\doi{10.1103/PhysRevLett.100.013001}.
\bibitem[{Royer(1977)}]{parityWigner}
\bibinfo{author}{A.~Royer}, \bibinfo{journal}{Phys. Rev. A}
  \bibinfo{volume}{15} (\bibinfo{year}{1977}) \bibinfo{pages}{449--450}.
  \URLprefix \url{http://link.aps.org/doi/10.1103/PhysRevA.15.449}.
  \DOIprefix\doi{10.1103/PhysRevA.15.449}.
\bibitem[{Bodenhausen et~al.(1976)Bodenhausen, Freeman, and Morris}]{DANTE1}
\bibinfo{author}{G.~Bodenhausen}, \bibinfo{author}{R.~Freeman},
  \bibinfo{author}{G.~A. Morris}, \bibinfo{journal}{J. Magn. Reson.}
  \bibinfo{volume}{23} (\bibinfo{year}{1976}) \bibinfo{pages}{171--5}.
  \DOIprefix\doi{10.1016/0022-2364(76)90150-5}.
\bibitem[{Morris and Freeman(1978)}]{DANTE2}
\bibinfo{author}{G.~A. Morris}, \bibinfo{author}{R.~Freeman},
  \bibinfo{journal}{J. Magn. Reson.} \bibinfo{volume}{29}
  (\bibinfo{year}{1978}) \bibinfo{pages}{433 -- 462}. \URLprefix
  \url{http://www.sciencedirect.com/science/article/pii/0022236478900033}.
  \DOIprefix\doi{10.1016/0022-2364(78)90003-3}.
\bibitem[{Ferretti and Ernst(1976)}]{Chirped1}
\bibinfo{author}{J.~A. Ferretti}, \bibinfo{author}{R.~R. Ernst},
  \bibinfo{journal}{J. Chem. Phys.} \bibinfo{volume}{65} (\bibinfo{year}{1976})
  \bibinfo{pages}{4283--4293}. \URLprefix
  \url{http://scitation.aip.org/content/aip/journal/jcp/65/10/10.1063/1.432837}.
  \DOIprefix\doi{http://dx.doi.org/10.1063/1.432837}.
\bibitem[{Jeschke and Schweiger(1995)}]{Chirped2}
\bibinfo{author}{G.~Jeschke}, \bibinfo{author}{A.~Schweiger},
  \bibinfo{journal}{J. Chem. Phys.} \bibinfo{volume}{103}
  (\bibinfo{year}{1995}) \bibinfo{pages}{8329--8337}. \URLprefix
  \url{http://scitation.aip.org/content/aip/journal/jcp/103/19/10.1063/1.470145}.
  \DOIprefix\doi{http://dx.doi.org/10.1063/1.470145}.
\bibitem[{Niemeyer et~al.(2013)Niemeyer, Shim, Zhang, Suter, Taniguchi, Teraji,
  Abe, Onoda, Yamamoto, Ohshima, Isoya, and Jelezko}]{Chirped3}
\bibinfo{author}{I.~Niemeyer}, \bibinfo{author}{J.~H. Shim},
  \bibinfo{author}{J.~Zhang}, \bibinfo{author}{D.~Suter},
  \bibinfo{author}{T.~Taniguchi}, \bibinfo{author}{T.~Teraji},
  \bibinfo{author}{H.~Abe}, \bibinfo{author}{S.~Onoda},
  \bibinfo{author}{T.~Yamamoto}, \bibinfo{author}{T.~Ohshima},
  \bibinfo{author}{J.~Isoya}, \bibinfo{author}{F.~Jelezko},
  \bibinfo{journal}{New J. Phys.} \bibinfo{volume}{15} (\bibinfo{year}{2013})
  \bibinfo{pages}{033027}. \URLprefix
  \url{http://stacks.iop.org/1367-2630/15/i=3/a=033027}.
\bibitem[{Tann{\'u}s and Garwood(1997)}]{Tannus1}
\bibinfo{author}{A.~Tann{\'u}s}, \bibinfo{author}{M.~Garwood},
  \bibinfo{journal}{NMR Biomed.} \bibinfo{volume}{10} (\bibinfo{year}{1997})
  \bibinfo{pages}{423--434}. \URLprefix
  \url{http://dx.doi.org/10.1002/(SICI)1099-1492(199712)10:8<423::AID-NBM488>3.0.CO;2-X}.
\bibitem[{Tann{\'u}s and Garwood(1996)}]{Tannus2}
\bibinfo{author}{A.~Tann{\'u}s}, \bibinfo{author}{M.~Garwood},
  \bibinfo{journal}{J. Magn. Reson. A} \bibinfo{volume}{120}
  (\bibinfo{year}{1996}) \bibinfo{pages}{133--137}.
  \DOIprefix\doi{10.1006/jmra.1996.0110}.
\bibitem[{Silver et~al.(1984)Silver, Joseph, and Hoult}]{HS1}
\bibinfo{author}{M.~Silver}, \bibinfo{author}{R.~Joseph},
  \bibinfo{author}{D.~Hoult}, \bibinfo{journal}{J. Magn. Reson.}
  \bibinfo{volume}{59} (\bibinfo{year}{1984}) \bibinfo{pages}{347 -- 351}.
  \URLprefix
  \url{http://www.sciencedirect.com/science/article/pii/0022236484901811}.
  \DOIprefix\doi{http://dx.doi.org/10.1016/0022-2364(84)90181-1}.
\bibitem[{Baum et~al.(1985)Baum, Tycko, and Pines}]{HS2}
\bibinfo{author}{J.~Baum}, \bibinfo{author}{R.~Tycko},
  \bibinfo{author}{A.~Pines}, \bibinfo{journal}{Phys. Rev. A}
  \bibinfo{volume}{32} (\bibinfo{year}{1985}) \bibinfo{pages}{3435--3447}.
  \URLprefix \url{http://link.aps.org/doi/10.1103/PhysRevA.32.3435}.
  \DOIprefix\doi{10.1103/PhysRevA.32.3435}.
\bibitem[{Gebert et~al.(2016)Gebert, Wan, Wolf, Heip, and Schmidt}]{Gebert2016}
\bibinfo{author}{F.~Gebert}, \bibinfo{author}{Y.~Wan},
  \bibinfo{author}{F.~Wolf}, \bibinfo{author}{J.~C. Heip},
  \bibinfo{author}{P.~O. Schmidt}, \bibinfo{journal}{New J. Phys.}
  \bibinfo{volume}{18} (\bibinfo{year}{2016}) \bibinfo{pages}{013037}.
  \URLprefix \url{http://stacks.iop.org/1367-2630/18/i=1/a=013037}.
\bibitem[{Watanabe et~al.(2011)Watanabe, Nomura, Toyoda, and
  Urabe}]{Watanabe2011}
\bibinfo{author}{T.~Watanabe}, \bibinfo{author}{S.~Nomura},
  \bibinfo{author}{K.~Toyoda}, \bibinfo{author}{S.~Urabe},
  \bibinfo{journal}{Phys. Rev. A} \bibinfo{volume}{84} (\bibinfo{year}{2011})
  \bibinfo{pages}{033412}. \URLprefix
  \url{http://link.aps.org/doi/10.1103/PhysRevA.84.033412}.
  \DOIprefix\doi{10.1103/PhysRevA.84.033412}.
\bibitem[{Linington et~al.(2008)Linington, Ivanov, Vitanov, and
  Plenio}]{Linington2008}
\bibinfo{author}{I.~E. Linington}, \bibinfo{author}{P.~A. Ivanov},
  \bibinfo{author}{N.~V. Vitanov}, \bibinfo{author}{M.~B. Plenio},
  \bibinfo{journal}{Phys. Rev. A} \bibinfo{volume}{77} (\bibinfo{year}{2008})
  \bibinfo{pages}{063837}. \URLprefix
  \url{http://link.aps.org/doi/10.1103/PhysRevA.77.063837}.
  \DOIprefix\doi{10.1103/PhysRevA.77.063837}.
\bibitem[{Toyoda et~al.(2011)Toyoda, Watanabe, Kimura, Nomura, Haze, and
  Urabe}]{Toyoda2011}
\bibinfo{author}{K.~Toyoda}, \bibinfo{author}{T.~Watanabe},
  \bibinfo{author}{T.~Kimura}, \bibinfo{author}{S.~Nomura},
  \bibinfo{author}{S.~Haze}, \bibinfo{author}{S.~Urabe},
  \bibinfo{journal}{Phys. Rev. A} \bibinfo{volume}{83} (\bibinfo{year}{2011})
  \bibinfo{pages}{022315}. \URLprefix
  \url{http://link.aps.org/doi/10.1103/PhysRevA.83.022315}.
  \DOIprefix\doi{10.1103/PhysRevA.83.022315}.

\end{thebibliography}

\end{document}